\documentclass[preprint, secnumarabic, amssymb, amsmath, aps, prd, nofootinbib]{revtex4-1}

\newcommand{\bra}[1]{\left\langle #1 \right|}
\newcommand{\ket}[1]{\left| #1 \right\rangle}

\newcommand{\ul}[1]{\underline{#1}}

\newcommand{\ubar}[1]{\overline{U}_{#1}}
\newcommand{\Tr}{\mathrm{Tr}}
\newcommand{\braket}[2]{\left\langle #1 \right| \left. #2 \right\rangle}

\newcommand{\half}{\frac{1}{2}}
\newcommand{\thalf}{\tfrac{1}{2}}
\newcommand{\ord}[1]{\mathcal{O}\left( #1 \right)}
\newcommand{\Jpsi}{J / \psi}

\newcommand{\gPM}{g^{+-}}

\DeclareMathOperator*{\SumInt}{%
\mathchoice%
  {\ooalign{$\displaystyle\sum$\cr\hidewidth$\displaystyle\int$\hidewidth\cr}}
  {\ooalign{\raisebox{.14\height}{\scalebox{.7}{$\textstyle\sum$}}\cr\hidewidth$\textstyle\int$\hidewidth\cr}}
  {\ooalign{\raisebox{.2\height}{\scalebox{.6}{$\scriptstyle\sum$}}\cr$\scriptstyle\int$\cr}}
  {\ooalign{\raisebox{.2\height}{\scalebox{.6}{$\scriptstyle\sum$}}\cr$\scriptstyle\int$\cr}}
}

\newcommand{\ben}{\begin{displaymath}}
\newcommand{\een}{\end{displaymath}}
\newcommand{\be}{\begin{equation}}
\newcommand{\ee}{\end{equation}}
\newcommand{\bea}{\begin{eqnarray}}
\newcommand{\eea}{\end{eqnarray}}

\usepackage{graphics, graphicx,slashed}
\usepackage{color}
\begin{document}

\preprint{NT@UW-15-16}
\title{Probing short-range nucleon-nucleon interactions with an Electron-Ion Collider}

\author{Gerald A. Miller}
	\email[Email: ]{miller@phys.washington.edu }
	\affiliation{Department of Physics, University of Washington, Seattle, WA 98195-1560, USA}
\author{Matthew D. Sievert}
	\email[Email: ]{msievert@bnl.gov}
	\affiliation{Bldg. 510A, Physics Department, Brookhaven National Laboratory, Upton, NY 11973, USA}
\author{Raju Venugopalan}
	\email[Email: ]{rajuv@bnl.gov}
	\affiliation{Bldg. 510A, Physics Department, Brookhaven National Laboratory, Upton, NY 11973, USA}
\affiliation{Institut f\"{u}r Theoretische Physik, Universit\"{a}t Heidelberg, Philosophenweg 16, 69120 Heidelberg, Germany
}
\date{\today}%

\begin{abstract}

We derive the cross-section for exclusive vector meson production in high energy deeply inelastic scattering off a deuteron target 
that disintegrates into a proton and a neutron carrying large relative momentum in the final state.  This cross-section can be expressed in terms of a novel gluon Transition Generalized Parton Distribution (T-GPD); the hard scale in the final state makes the T-GPD sensitive to the short distance nucleon-nucleon interaction. We perform a toy model computation of this process in a perturbative framework and discuss the time scales that allow the separation of initial and final state dynamics in the T-GPD. We outline the more general computation based on the factorization suggested by the toy computation: in particular, we discuss the relative role of ``point-like" and ``geometric" Fock configurations that control the parton dynamics of short range nucleon-nucleon scattering. With the aid of exclusive $J/\Psi$ production data at HERA, as well as elastic nucleon-nucleon cross-sections, we estimate rates for exclusive deuteron photo-disintegration at a future Electron-Ion Collider (EIC). Our results, obtained using conservative estimates of EIC integrated luminosities,  suggest that center-of-mass energies $s_{NN}\sim 12$ GeV$^2$ of the neutron-proton subsystem can be accessed. We argue that the high energies of the EIC can address outstanding dynamical questions regarding the short-range quark-gluon structure of nuclear forces by  providing clean gluon probes of such ``knockout" exclusive reactions in light and heavy nuclei.

\end{abstract}


\maketitle



\section{Introduction}
\label{sec:Intro}

The finite range of the strong force ensures that the nucleon-nucleon interaction plays a vital role in the structure of atomic nuclei. There is a spatial separation of the nucleon-nucleon potential into three parts which is often captured in distinct theoretical treatments. The long range part of the nuclear force is described by pion exchange, and is well understood within the framework of effective field theories. At shorter distances, two pion exchange, tensor interactions, and vector meson exchange contributions become important, which are harder to capture in the framework of effective field theories. At short distances, the nucleon-nucleon interaction has a strong repulsive core, which is essential for the stability of matter. There has been much recent progress from lattice gauge theory in first principles simulations of the nucleon-nucleon potential~\cite{Savage:2013wwa}. Because  the virtuality of the exchanged particles at short distances is large, it is natural to consider to what extent the short distance contributions to nucleon-nucleon scattering can be described directly in terms of the  fundamental quark and gluon degrees of freedom in Quantum Chromodynamics (QCD). 

There has been a significant amount of work in trying to understand short-range contributions to nucleon-nucleon (NN) collisions using quark and gluon degrees of freedom. For a nice recent review, we refer the reader to Ref.~\cite{Sargsian:2014bwa} and references therein. The interest in short range nucleon-nucleon correlations has been rekindled by the discovery at Jefferson Lab of the strong dominance of short range proton-neutron correlations over neutron-neutron and proton-proton correlations~\cite{Piasetzky:2006ai}  based on expectations from quark-gluon dynamics anticipated over two decades ago~\cite{Frankfurt:1993sp}. Remarkably the systematics of such short range correlations appear to be empirically correlated with nuclear modifications termed the EMC effect that were first observed in deeply inelastic scattering (DIS) experiments by the European Muon Collaboration~\cite{Hen:2013oha,Higinbotham:2013hta,Hen:2014vua,Weinstein:2010rt}. It has also been argued that such short range nuclear forces, and their parton interpretation in particular, have significant implications for the neutron star equation of state~\cite{Yamamoto:2015lwa,Yamamoto:2013ada}.

In this paper, we will address the possibility that a high luminosity, high energy, electron-ion collider (EIC) can provide novel information on short range nucleon-nucleon interactions. The construction of such a machine is a prominent recommendation of the Nuclear Physics community in their recent Long Range Plan~\cite{Geesaman:2015fha}. Detailed discussions of the science case for such a machine in either the US or at CERN can be found in \cite{Boer:2011fh,Accardi:2012qut,AbelleiraFernandez:2012cc}; a brief overview can be found in \cite{Venugopalan:2015jsa}. In all EIC proposals, Bjorken $x < 10^{-2}$ and momentum transfer squared $Q^2 \geq 10$ GeV$^2$ will be achieved for light and heavy nuclei. This high energy kinematics allows for a clean separation of the current fragmentation region of the virtual photon projectile from the fragmentation region of the nuclear target. In particular, in diffractive processes where no net color charge is exchanged between the current and nuclear fragmentation regions, the quark and gluon degrees of freedom from the former can cleanly probe relatively low energy nucleon-nucleon interactions in the latter. The precise kinematics of such experiments, the ability to vary the size and nature of the probe, and the availability of a range of nuclear targets (including polarized light nuclei) have the potential to open up a new window into the parton structure of nuclear forces. It is therefore a problem of considerable interest to estimate whether the rates necessary for a comprehensive study of the nuclear fragmentation region can be achieved with the luminosities projected for an EIC. The peak luminosities are estimated to be a hundred to a thousand times larger than the peak luminosity for electron-proton DIS at the HERA collider. 

We will consider here, for specificity, the exclusive process $e+D \longrightarrow e + J/\Psi + n +p$, namely, the electroproduction of $J/\Psi$ mesons in coincidence with a proton and a neutron produced with relative transverse momenta of a GeV or larger from the disintegration of the struck deuteron. The exclusive electroproduction of heavy quarkonia has long been known to be a sensitive probe of the QCD degrees of freedom in a hadronic target. The hard scale introduced by the mass of a heavy vector meson like the $\Jpsi$ ensures that the process is sensitive to short-distance physics, and is therefore perturbatively calculable \cite{Collins:1996fb, Frankfurt:1997fj,Ivanov:2004vd}.  Real or virtual photon-hadron scattering in Regge kinematics, in which the center-of-mass energy $\sqrt{s}$ is much larger than any other kinematic scale, is characterized by the fluctuation of the photon into a quark-antiquark dipole which then scatters from the target by exchanging gluons \cite{Kopeliovich:1981pz, Bertsch:1981py, Mueller:1989st, Nikolaev:1990ja}.  Thus the exclusive production of heavy vector mesons is directly sensitive to the gluon distribution in the target state \cite{Brodsky:1994kf, Frankfurt:1995jw, Frankfurt:1997fj}.  When the diffractive gluon exchange transfers net momentum to the hadronic target, the process couples the dipole to the generalized parton distribution (GPD) $H^g$ of gluons (see for instance the review \cite{Diehl:2003ny}); when the target is a proton, in the forward limit, the dipole couples to the integrated gluon distribution $x G(x,Q^2)$ \cite{Brodsky:1994kf, Frankfurt:1995jw, Frankfurt:1997fj}.

The above-mentioned ``dipole picture'' works because (in Regge kinematics) there is a separation of multiple scales between the physics of the projectile and the physics of the deuteron target.  It is important that the virtual momentum squared of the photon $Q^2$, or equivalently, the invariant mass of the vector meson that the quark-antiquark pair hadronizes into, be very large compared to the intrinsic QCD scale. Then the transverse dipole size is small enough that two-gluon exchange in a color-singlet ``pomeron" configuration is the dominant process in the scattering off the deuteron target.  Furthermore, in a co-moving frame with the dipole, the time scale over which this exchange occurs is nearly instantaneous because the target is Lorentz-contracted to a scale which is suppressed by the DIS center-of-mass energy. As a consequence of these short times, the two gluons exchanged with the target can probe a number of Fock configurations of the deuteron bound state whose lifetimes are relatively large. What is novel about the particular kinematics we consider is that the outgoing proton and neutron have large relative transverse momenta and are therefore sensitive to the short-range components of these Fock configurations. 

The interaction of the two gluons with the deuteron target can be expressed most generally in terms of a novel gluon Transition Generalized Parton Distribution (T-GPD).  This object is the expectation value of  an operator corresponding to the product of the color field strengths of the two gluons\footnote{As we will elaborate, these are dressed with ``gauge links" to ensure the operator is gauge invariant.} acting on the Fock state of the deuteron on one side and the product of the proton and neutron Fock states on the other. Because the probability that the deuteron can be found in a two-nucleon configuration is significant on the time scale of the gluon exchange, in this case the T-GPD reduces to the two-gluon exchange operator sandwiched between the product of the nucleon states and is therefore much closer in spirit to the usual GPD distribution~\cite{Diehl:2003ny}. A significant difference is that the former now depends on two momentum transfer scales; one momentum scale comes from the projectile ($T = \Delta^2$), the square of the four momentum transferred to the vector meson, and the new scale $t = (p_1^\prime - p)^2$, the square of the four-momentum transferred to the struck nucleon, is contained wholly within the target.  The T-GPD therefore tells us about the short distance structure of the target when $t$ is large.

When the final-state proton and neutron have low relative transverse momentum (less than or of the order of the typical Fermi momentum they possess in the deuteron bound state) the leading-order process corresponds to the two gluons being exchanged with either the proton or the neutron; the other nucleon acts a spectator, albeit it may exchange soft gluons with the other nucleon. But if one detects the final-state nucleons with \textit{larger} $p_T$ than the momentum transfer $T$ from the projectile dipole, $p_T^2 \gg |T|$, this extra transverse momentum must have originated from the interactions between the nucleons themselves.  If this momentum is perturbatively large, $p_T^2 \gg \Lambda_{QCD}^2$ (where $\Lambda_{\rm QCD}$ is the intrinsic QCD scale),  the nucleon-nucleon interaction should be mediated by quark and gluon degrees of freedom.  

Despite the high energy $\sqrt{s}$ of the $\gamma D$ system, the center-of-mass energy of the interacting nucleons is much lower: $s_{NN} \sim 4 p_T^2$.  In such events, the disintegration of the deuteron is sensitive to the nucleon-nucleon scattering amplitude at these lower energy scales.  One is therefore using very high energy DIS to probe nucleon-nucleon interactions at much lower energies, 
$s_{NN} \ll s$.

A further interesting possibility suggested by diffractive DIS at an EIC is the potential to distinguish between differing parton models of the elastic nucleon-nucleon scattering as a function of $s_{NN}$. One such model is the ``geometric'' or ``independent quark scattering model" whereby collinear quarks from one nucleon each individually exchange a gluon with a quark from the other nucleon. Another is a ``point-like scattering model,'' in which all the valence quarks of the colliding nucleons participate in a single short-distance hard scattering.  From perturbative power counting, the former ``Landshoff" mechanism~\cite{Landshoff:1974ew} predicts cross-sections which fall as $s_{NN}^{-8}$, while the latter ``quark counting"  mechanism~\cite{Brodsky:1974vy, Matveev:1973ra} predicts that the cross-sections should fall as $s_{NN}^{-10}$. This latter dependence is what is seen in data, albeit significant  differences are seen at low values of $s_{NN}$. 

The energy dependence of the nucleon-nucleon interaction, as extracted in diffractive DIS, could add a different twist to this picture; due to the external gluon probe, the power counting may differ from that extracted in nucleon-nucleon elastic scattering. For instance, there are both color-singlet as well as color-octet Landshoff exchanges feasible in the deuteron disintegration process. Further insight into such parton-based mechanisms of short-range correlations is highly likely at an EIC because, as we will show, the rates estimated for the exclusive deuteron photo-disintegration are very sensitive to the energy dependence of the nucleon-nucleon cross-section.  The potential of the EIC to polarize light nuclei will add equally important insight to the study of spin dependent nucleon-nucleon interactions -- we leave such a study for future work.

The deuteron can also in principle fluctuate into exotic  ``hidden color" bound color octet-color octet configurations which the instantaneous
gluon exchanges with the projectile are capable of resolving~\cite{Harvey:1988nk,Brodsky:1983vf}.  The hidden color nucleons can be liberated if struck by hard back-to-back gluons from the projectile dipole, projecting them onto back-to-back nucleons in the final state with a characteristic transverse momentum dependence. While a very intriguing possibility, this mechanism to directly access hidden color configurations breaks the factorization between projectile and target. In a parallel investigation, we nevertheless estimated the rates for such hidden color states; we find these to be prohibitively small even at the high luminosities of EIC. It is conceivable though that other similar channels to access such states may exist. 

This paper is organized as follows. In Section~\ref{sec:kin}, we will begin with an outline of the kinematics of an EIC and why they may be favorable to elucidate key aspects of the short range structure of nuclear forces. We will then discuss quarkonium production in the kinematics of high energy DIS. As a warmup, we will first consider exclusive $J/\Psi$ production off the proton before generalizing our results to the deuteron case. In particular, we will obtain an explicit expression for the cross-section for the exclusive electroproduction of vector mesons off a deuteron target accompanied by the breakup of the deuteron.  When either the photon virtuality or vector meson mass is a hard scale, the result is proportional to the gluon T-GPD. We will compare and contrast this object with the GPD probed in the proton case and argue that, if the proton and neutron fly off back-to-back, each with $p_T\gg \Lambda_{\rm QCD}$, the T-GPD extracted from experiments at the EIC will contain novel information about short range nucleon-nucleon interactions. 

In Section~\ref{sec:pQCD}, we will discuss the simplifications of this T-GPD that occur when there is a large relative momentum between the proton and the neutron. In this case, a factorization appears to occur between the nonperturbative wavefunction of the deuteron and the final state interactions between the neutron and the proton.  We shall perform an explicit ``toy model"  computation (with the proton and the neutron replaced by valence quarks) that explicitly illustrates this factorization and establishes a baseline for semi-quantitative estimates for the photo-disintegration cross-section. We will next discuss how the toy model computation generalizes to the realistic case. The issues are strongly related to the extensive literature on large angle nucleon-nucleon scattering and the insight they provide on the parton configurations contributing to short range nuclear forces.  We will exploit these lessons to make an ansatz for the cross-section for exclusive $J/\Psi$ photo-disintegration of the deuteron: the result can be expressed as the product of the deuteron wavefunction times the cross-section for exclusive photo-production of $J/\Psi$ mesons off the proton times the elastic nucleon-nucleon cross-section.  

In Section~\ref{sec:rates}, we will use the results of the previous section to estimate the rates for this process at an EIC. We find, for conservative estimates of the EIC luminosity, that rates comparable to those used to extract the precise HERA data on exclusive photo-production of $J/\Psi$ can be achieved for center-of-mass energies of the proton-nucleon subsystem of up to $s_{NN}\sim 12$ GeV$^2$. The extension of the cross-section out to significantly higher $s_{NN}$ with increasing luminosity will be be challenging if the energy dependence of the cross-section is close to those measured in nucleon-nucleon scattering. Nevertheless, the range covered should be sufficient to provide considerable insight into the transition from hadron to parton degrees of freedom in the description of short range nuclear forces. We conclude with a discussion of open issues and prospects for future work.  Appendix~\ref{sec:JerryWF} contains some details of the deuteron  wavefunction employed in our estimates.

\section{Exclusive $J/\Psi$ production in high-energy DIS}
\label{sec:kin}


We will begin our discussion in Section~\ref{sec:EICgen}  by considering the relevant parameters at an EIC for the process of interest.  We will then in Section~\ref{sec:protonkin} derive the well-known expression for exclusive $J/\Psi$ production in high-energy DIS off the proton. In light-front perturbation theory (LFPT) the natural separation of time scales is made explicit, whereby the scattering amplitude can be factorized into the amplitude of the virtual or real photon to fluctuate into a quark-antiquark pair which is relatively long lived, and the amplitude of the dipole to scatter off the target nucleus before recombining into the heavy quarkonium state. The latter time scale is much longer than the time scale of the interaction of the  elastic scattering of the quark-antiquark pair off the target. We will show that the forward matrix element of the quark-antiquark interaction with the target is proportional to a generalized parton distribution (GPD). 

In Section~\ref{sec:IntroDeut}, we will extend our analysis to the problem of interest: the exclusive production of $J/\Psi$ in DIS off the deuteron, accompanied by the proton and neutron that are produced back-to-back. We will compute the cross-section and show that, in analogy to the proton case, it is proportional to a novel gluon Transition Generalized Parton Distribution (T-GPD). As with the discussion of the separation of the time scales in the dipole projectile, we will argue that a similar factorization of time scales occurs in the T-GPD of the target when the relative transverse momentum of the final state proton and the neutron is large. In this case, the T-GPD can be factorized into the deuteron wavefunction convoluted with a matrix element that is sensitive to the parton structure of the short-range nuclear force.

\subsection{Insights into Nuclear Structure from an EIC}
\label{sec:EICgen}

At a future electron-ion collider (EIC) project in the US \cite{Accardi:2012qut}, high-energy electrons will collide with nuclei at center-of-mass energies around $\sqrt{s} \sim 100$~GeV/nucleon; significantly higher energies will be feasible with the proposed LHeC collider~\cite{AbelleiraFernandez:2012cc}. In this work, for specificity, we will make estimates for EIC alone; these can easily be extended to studies with LHeC kinematics.   High rates of $\Jpsi$ production occur when the virtuality $Q^2$ becomes small, approaching the photoproduction limit as $Q^2 \rightarrow 0$.  In this regime, the EIC will access values of $x$ below $\sim 10^{-3}$, well into the domain of Regge kinematics $x \ll 1$.  

The diffractive $\Jpsi$ photoproduction cross-section is a steeply falling function of the exchanged momentum $|T|$, so it is advantageous to look for events in which the $\Jpsi$ is produced with fairly low transverse momentum (relative to the photon) of a few hundred MeV or so.  Events of this class have already been observed on proton targets at HERA \cite{Chekanov:2002xi} with adequate statistics. Since an EIC would have a luminosity  orders of magnitude higher ($\sim 10^{33-34} \, cm^{-2} s^{-1}$ compared to the peak luminosity $\sim 10^{31} \, cm^{-2} s^{-1}$ at HERA), low-$p_T$ diffractive $\Jpsi$ photoproduction should be relatively easy to observe at an EIC.  

The high-$p_T$ disintegration of the deuteron in the process of interest here, $\gamma + D \rightarrow \Jpsi + p +n$, only occurs in a subset of the diffractive events.  To isolate it, an additional cut on the $p_T$ of the detected nucleons must be imposed: the nucleons should emerge nearly back-to-back, with $p_T$ much larger than the $T$-channel momentum seen in the $\Jpsi$.  In this regime, an additional nucleon-nucleon $(NN)$ scattering must have occurred, with a relative center-of-mass energy squared set by the transverse momentum $s_{NN} \sim 4 p_T^2 $.  If the nucleon $p_T$ is larger than a GeV or so, then the $NN$ rescattering takes place over ``perturbatively" short distances where it should be mediated by quark and gluon degrees of freedom.  
These events are rarer than the diffractive $\Jpsi$ photoproduction events observed at HERA. However because of the increase in luminosity at an EIC, they may be measured with reasonable statistics.  We will explore this question quantitatively in Sec.~\ref{sec:rates}.

The study of the final state proposed here utilizes the interplay of several scales - hard and soft - to access the $NN$ interaction in a novel way:  
\begin{align} \label{e:kinnum}
  & \underset{( \, \sim \, 100~GeV)^2}{s} \, \gg \, \underset{(\, \sim \, 1~GeV)^2}{p_T^2}  \, \gg \, \underset{(\, \sim \, \mathrm{few}~100~MeV)^2}{|T|}
  \notag \\ &\hspace{1.2cm}
  \underset{(\, \sim \, 3~GeV)^2}{M_J^2} \, \gg \, \underset{(\, \sim \, 200\, MeV)^2}{\Lambda_{QCD}^2}.
\end{align}
The photon-deuteron center-of-mass energy $\sqrt{s}$ is the hardest scale in the process. It ensures that the projectile-deuteron interaction is effectively instantaneous, providing a snapshot of the deuteron wave function.  The mass $M_J$ of the $\Jpsi$ is another hard scale which makes the instantaneous diffractive exchange calculable within perturbative QCD (pQCD).  The invariant momentum transfer $|T|$ between the projectile and the deuteron is a soft scale which maximizes the diffractive cross-section.  The nucleon recoil momentum $p_T$ is an intermediate scale; it is hard enough to guarantee that the nucleons emerge back-to-back (and that the rescattering is perturbative), yet still small relative to the center-of-mass energy.  

Thus the EIC, a machine designed to access high-energy nuclear physics at energies on the order of $ \sqrt{s}/A \sim 100$ GeV, can also be used to study physics relevant for nuclear structure at much lower energies $\sqrt{s_{NN}} \sim$ few GeV.  This convergence of nuclear physics from both sides of the energy spectrum provides an opportunity to learn unique information about the $NN$ potential at short distances and is wholly complementary to the conventional approaches which measure $NN$ scattering directly. We only consider unpolarized scattering here; similar exclusive processes off polarized light nuclei will offer additional unique opportunities to explore the spin-dependent nature of such short range correlations. 


\subsection{Dipole picture of heavy quarkonium production in high energy DIS off the proton}
\label{sec:protonkin}

Photon-hadron scattering in Regge kinematics (small-$x$) is intuitively described in terms of the infinite-momentum scattering formalism of \cite{Kogut:1969xa, Bjorken:1970ah}.  For an ultrarelativistic system moving along the $x^+$ axis
\footnote{Light-front coordinates are defined by  $x^\pm \equiv \sqrt{\frac{\gPM}{2}} \left( x^0 \pm x^3 \right)$, where $\gPM = 1$ and $\gPM = 2$ are two common choices of the light-front metric.  Here we will use $\gPM = 1$.}
, one quantizes the theory at a fixed $x^+$.  The resulting light-front perturbation theory (LFPT) \cite{Kogut:1969xa, Bjorken:1970ah, Lepage:1980fj, Brodsky:1997de, Kovchegov:2012mbw} corresponds to ``old-fashioned'' time-ordered perturbation theory where $x^+$ plays the role of time and the light-front momentum operator $p^- = \mathcal{H}^-$ plays the role of the Hamiltonian.  In the infinite momentum limit, the interaction of the high energy projectile with the target is nearly instantaneous ($\delta x^+_{Int} \propto \tfrac{1}{\sqrt{s}}$), providing a ``snapshot'' of the configuration of the target occurring at a light front time we define to be $x^+ \equiv 0^+$.  Then the scattering matrix of the high-energy projectile consists of ``time'' evolution from $x^+ \! = \! -\infty^+$ to $0^+$, an instantaneous eikonal scattering in the field $A^{-a}$ of the target and the ``time'' evolution from $x^+ \! = 0^+$ to $+\infty^+$~\cite{ Bjorken:1970ah}:
\begin{align}
S_{fi} = \langle f |{\cal P} \exp\left( -i\int dz^+ {\cal H}_I^- (z^+)\right)|i \rangle  \,,
\label{e:full-S}
\end{align}
where ${\cal H}_{I}^- = {\cal H}^- - {\cal H}_0^-$ is the interaction-picture Hamiltonian, with ${\cal H}_0^-$ the free-particle Hamiltonian, and 
\begin{align}
|i\rangle = |e^-\rangle \otimes  |D\rangle \:\: ; \:\: |f\rangle = |J/\Psi\rangle  \ket{e^-} \otimes \ket{n} \ket{p} \, .
\end{align}
Defining $S_{fi} = \bra{n} \bra{p} \: {\tilde S}_{fi} \: \ket{D}$, 
one can express the ``projectile" S-matrix as 
\begin{align} \label{e:BKS1}
 {\tilde S}_{fi} &= \bra{\Jpsi} \mathcal{U}[+\infty^+ , 0^+] \: \Tr_C \mathcal{P} \exp\left[ - i \int d^4 x \, j^+ (0^+, x^-, \ul{x}) \, 
  A^{-a} (x^+, 0^-, \ul{x}) T^a \right]
  \notag \\ & \hspace{2cm} \times
  \mathcal{U}[0^+ , -\infty^+] \ket{\gamma} \, ,
\end{align}
where $\Tr_C$ stands for a trace over color matrices, $T^a$ are the generators of $SU(N_c)$ in the fundamental representation, and $\ul{x} \equiv (x_\bot^1 , x_\bot^2)$ is a two-vector in the transverse plane.  Also  $j^+$ is the eikonal current of the high-energy quanta at the collision time $x^+ = 0^+$, and the gluon fields $A^{-a}$ of the target are path-ordered $\mathcal{P}$ along the $x^+$-direction.  The ``time'' evolution operator in the interaction picture is
\begin{align} \label{e:timeev}
     \mathcal{U}[x_f^+, x_i^+] \equiv \exp\left[i \mathcal{H}_0^- x_f^+ \right] \exp\left[-i \mathcal{H}^- (x_f^+ - x_i^+)\right]
     \exp\left[-i \mathcal{H}_0^- x_i^+ \right],
\end{align}
and we emphasize that the gluon fields $A^{- a}$, and $\tilde S$ itself, are operators which are evaluated between the target states.  The photon $\ket{\gamma}$ can fluctuate into one of many Fock states $\ket{\mathbb{X}}$ before it scatters on the target,
\begin{align}
  &\tilde S = \SumInt_{\mathbb{X}} \SumInt_{\mathbb{X}'} \bra{\Jpsi} \mathcal{U}[+\infty^+ , 0^+] \ket{\mathbb{X}'} \:\:
  \bra{\mathbb{X}} \mathcal{U}[0^+ , -\infty^+] \ket{\gamma}
  \notag \\ & \hspace{2cm} \times 
  \bra{\mathbb{X}'} \Tr_C \mathcal{P} \exp\left[ - i \int d^4 x \, j^+ (0^+, x^-, \ul{x}) \, A^{-a} (x^+, 0^-, \ul{x}) T^a\right] 
  \ket{\mathbb{X}} \, .
\end{align}
Here $\SumInt_{\mathbb{X}}$ denotes a complete sum over Fock states and integrals over their phase spaces.  

The leading contribution in QED is the fluctuation of the photon into a quark-antiquark dipole $\ket{q \bar q}$ which can scatter by gluon exchange; this light-front wave function (LFWF) is given by
\begin{align}
	\psi^{\gamma \rightarrow q \bar q} \equiv \bra{q \bar q} \mathcal{U} [0^+, -\infty^+] \ket{\gamma} \approx
         \frac{\bra{q \bar{q}} (\mathcal{H}_{I \, , \, QED}^-) \ket{\gamma}}
	{p^-_{\gamma} - p^-_q - p^-_{\bar{q}}} .
\end{align}
Truncating the Fock space $\ket{\mathbb{X}}$ at the quark dipole, we obtain for the ${\tilde S}$-matrix
\begin{align}
 {\tilde S}_{fi} &=  \int d\Omega_{q \bar q}  \, (\psi^{\Jpsi \rightarrow q \bar q})^* \, \psi^{\gamma \rightarrow q \bar q} 
 \notag \\ & \hspace{2cm} \times
 \bra{q \bar q} \Tr_C \mathcal{P} \exp\left[ - i  \int d^4 x \, j^+ (0^+, x^-, \ul{x})  A^{-a} (x^+, 0^-, \ul{x}) T^a\right] \ket{q \bar q} ,
\end{align} 
where we have similarly defined the LFWF of the $\Jpsi$ and $\int d\Omega_{q \bar q}$ represents an integral over the phase space of the dipole.

In ordinary time-ordered perturbation theory, the energy denominator $1/\Delta E \sim \Delta t$ is a direct measure of the lifetime $\Delta t$ of a virtual fluctuation, allowing a clear physical interpretation of the process.  The same is true of LFPT, with $\Delta x^+$ and $\Delta p^-$ playing analogous roles.  For example, for $\Jpsi$  electroproduction, there are three relevant time scales: the lifetime $\Delta x^+_{q \bar q}$ of the $q \bar q$ pair, the formation time $\Delta x^+_{J}$ of the $\Jpsi$, and the duration of the interaction $\Delta x^+_{Int}$.  Using LFPT to calculate the associated energy denominators, one readily notes a distinct separation of time scales \cite{Brodsky:1994kf}:
\begin{align}
  \frac{\Delta x^+_{q \bar q}}{p^+_{q \bar q}} \sim \ord{\frac{1}{m_c^2}} \hspace{0.75cm}
	\frac{\Delta x^+_{J}}{p^+_{J}} \sim \ord{\frac{1}{M_J^2}} \hspace{0.75cm}
	\frac{\Delta x^+_{Int}}{p^+_{tot}} \sim \ord{\frac{1}{s}} \,,
\end{align}
where $m_c$ is the mass of the charm quark and  the subscript $J$ refers to the $\Jpsi$, {\it e.g} $M_J$ is the mass of the $\Jpsi$. Since $\Delta x^+_{J} \, , \ \Delta x^+_{q \bar q} \gg \Delta x^+_{Int}$, the energy denominators in the $q \bar q$ scattering matrix element off the target become independent of the long-time dynamics in the photon and $\Jpsi$ wave functions, factorizing \cite{Ivanov:2004vd} into the on-shell dipole scattering amplitude $\Tr_C {\tilde M}^{q \bar q}$:
\begin{align}
  {\tilde M}^{\gamma \rightarrow \Jpsi} = \int d\Omega_{q \bar q} \: \: \left( \psi^{J \rightarrow q \bar q} \right)^* \, 
  \Tr_C {\tilde M}^{q \bar q} \:\:  \psi^{(\gamma \rightarrow q \bar q)} ,
\end{align} 
where we have subtracted the noninteracting term $\mathbf{1}$ from the ${\tilde S}$-matrix to form the scattering amplitude ${\tilde M}^{\gamma \rightarrow \Jpsi}$.

To determine the full S-matrix in Eq.~(\ref{e:full-S}),  one must fix the dipole scattering amplitude on the target.  As a warmup to DIS off the deuteron with disintegration into back-to-back nucleons in the final state, we will first compute explicitly the cross-section for elastic vector meson production on the proton, $\gamma^* (q) + p(p) \rightarrow V (q-\Delta) + p (p + \Delta)$. This process was measured at HERA and therefore will also be relevant in our discussion of rates for the extension to the deuteron case at the EIC. 

The center-of-mass energy (squared) of the photon-proton system is $s \equiv (p + q)^2$, and the invariant momentum transfer is $T \equiv \Delta^2$.  We will work in the photon-proton center-of-mass frame and choose the latter to move along the $\oplus$ axis with large longitudinal momentum $p^+$ and the photon to move in the opposite direction with large longitudinal momentum $q^-$ so that $s \approx 2 p^+ q^-$.  The incoming momenta of the proton and photon, respectively, are
\begin{align}
  p^\mu = \left( p^+ , \tfrac{m_N^2}{s} q^-, \ul{0} \right) \hspace{0.5cm} q^\mu &= \left( - x \, p^+ , q^- , \ul{0} \right) ,
\end{align}
where $x \approx \tfrac{Q^2}{s} \ll 1$.  The on-shell conditions for the vector meson and the proton fix the longitudinal components of the momentum transfer $\Delta^\mu$ to be
\begin{align} \label{e:Delkin}
  \Delta^+ = q^+ - (q-\Delta)^+ \approx 
  -  p^+ \equiv - x_{eff} \, p^+ \,\,\,;\,\,\,\Delta^- = p^{\prime -} - p^- \approx \left( \frac{p_T^{\prime 2}}{s} \right) q^- ,
\end{align}
where $x_{eff}=\left( \frac{Q^2 + M_V^2}{s} \right)$ and the outgoing proton momentum is $p' = p + \Delta$,  $M_V$ is the mass of the produced vector meson, and we have neglected $\Delta_T^2 \ll \max(Q^2 , M_V^2)$.  Further, $p_T^{\prime 2} = \Delta_T^2$ is the transverse momentum of the scattered proton. 

The production cross-section is then
\begin{align}
	&d\sigma = 
	\frac{1}{2 s} \frac{1}{4(2\pi)^2} \, \frac{d^2\Delta \, d\Delta^-}{q^- - \Delta^-} \, 
	\frac{d^2 p^\prime \, dp^{\prime +}}{p^{\prime +}} \, | M |^2 
	\notag \\ & \hspace{2cm} \times	
\delta (p^+ + \Delta^+ - p^{\prime +}) \delta (p^- + \Delta^- - p^{\prime -} ) 
        \delta^2 (\ul{\Delta} - \ul{p^\prime}) \, .
\end{align}
This gives $d \sigma \approx\frac{d^2 \Delta}{(2\pi)^2} \left| \frac{M}{2 s} \right|^2$; defining the invariant momentum transfer $T \equiv \Delta^2 \approx - \Delta_T^2$ and rescaled amplitude $A \equiv M / 2 s$ to write $d\sigma \approx \frac{d T}{4\pi} \, \frac{d \phi_\Delta}{2\pi} \, |A|^2$, 
where $\phi_\Delta$ is the azimuthal angle of $\ul{\Delta}$, we obtain
\begin{align}
\label{e:proton-cs}
     \frac{d\sigma^N}{dT} = \frac{1}{4\pi} \int \frac{d\phi_\Delta}{2\pi} \: | A |^2
\end{align}
for the cross-section for exclusive vector meson production on the nucleon.

As discussed previously, due to the separation of time scales in high energy Regge kinematics, manifest in LFPT, the amplitude $A$ can be decomposed into a convolution of the light front wave functions of the photon and vector meson with the on-shell scattering amplitude $A^{q \bar q N}$ for the scattering of the quark dipole on the nucleon as 
\begin{align} \label{e:dipfact}
  A (\ul{\Delta}) &= \int \frac{d^2 r \, dz}{4\pi z (1-z)} \,
  \left[ \sum_{\sigma \sigma'} \, \psi^{\gamma}_{\lambda \sigma \sigma'} (\ul{r}, z) 
  \, \left( \psi^V_{\lambda' \sigma \sigma'} (\ul{r}, z) \right)^*  \right]
  %
  \,  \Tr_C \: A^{q \bar q N} (\ul{r}, \ul{\Delta}).
\end{align}
Here $\ul{r}$ is the separation vector of the quark dipole, $z$ is the fraction of the photon's momentum carried by the quark, $\sigma (\sigma')$ are the spins of the quark (antiquark) in the dipole, $\lambda$ is the polarization of the photon, and $\lambda'$ is the polarization of the produced vector meson.  The notation $\Tr_C$ indicates a trace over the fundamental color representation of the dipole, and we have used the conventions of \cite{Kovchegov:2012mbw} for the normalization of the wave functions and phase space integrals.  Note that in the above decomposition, we  have  used the fact that the eikonal scattering of the quark-antiquark pair off the target is spin independent and helicity conserving.  We also remark that in the literature (for example, \cite{Rezaeian:2012ji}) it is common to assign a color factor $\sqrt{N_c}$ to the wave functions so that the dipole scattering amplitude enters with $\frac{1}{N_c} \Tr_C A^{q \bar q N}$.  We use a convention in which the color factor is contained entirely within the dipole scattering amplitude.

If a hard scale is present in the quark loop, then the quark dipole has small transverse size.  This can occur either if the photon has large virtuality $Q^2 \gg \Lambda_{QCD}^2$ (Deeply Virtual Meson Production), or if the vector meson has a heavy mass $M_V^2 \gg \Lambda_{QCD}^2$ (Heavy Exclusive Meson Production).  The typical transverse size of the dipole is set by the photon wave function to be $\frac{1}{r_T^2} \sim \langle k_T^2 \rangle \sim z (1-z) Q^2 + m_q^2$. 
If either $Q^2$ or the quark mass $m_q^2$ is large, then the dipole becomes perturbatively small.  For the case of large $Q^2$, special care must be taken in the limits $z \rightarrow 0, 1$ \cite{Kovchegov:1999kx, Kovchegov:2015zha}.  At lowest order, the dipole scatters on the
proton by exchanging two gluons, effectively measuring the strength of its gluon field.  For scattering on an unpolarized proton in these Regge kinematics, the dipole scattering amplitude is given by a trivial ``hard factor,'' times a nonperturbative gluon matrix element of the proton, the generalized parton distribution (GPD) $H^g (x,\xi, T)$ of the nucleon $(N)$:
\begin{align} \label{e:dipfact2}
  \Tr_C \, A^{q \bar q N} (\ul{r}, \ul{\Delta}) \approx \frac{\alpha_s \pi^2}{2} r_T^2 \, H^g_{(N)} (x_{eff}, 0, - \Delta_T^2),
\end{align}
where in the light-cone gauge $A^+ = 0$, the gluon GPD $H^g$ is given by the matrix element
\begin{align} \label{e:GPD1}
   H^g (x, 0, -\Delta_T^2) = \frac{1}{2\pi p^+} \int dr^- \, e^{i x p^+  r^-} \, \bra{p + \half\Delta} F^{+ i a}(-\thalf r) F^{+ i a} (+\thalf r)
   \ket{p - \half\Delta}.
\end{align}
The coefficient in \eqref{e:dipfact2} is fixed by a direct computation of both sides of the equation for the case of a quark target, where one has
\begin{align} \label{e:protondip1}
  \Tr_C \, A^{q \bar q N} (\ul{r}, \ul{\Delta}) \approx i \alpha_s^2 C_F
	\int d^2 b \: e^{i \ul{\Delta} \cdot \ul{b}} \: &\Big[
    \ln^2 \tfrac{1}{|b - z r|_T \Lambda} - 2 \ln \tfrac{1}{|b - z r|_T \Lambda} \ln \tfrac{1}{|b+(1-z)r|_T \Lambda} 
    \notag \\ &
    + \ln^2 \tfrac{1}{|b + (1-z) r|_T \Lambda} \Big].
\end{align}
In general, the field strength tensors in the operator \eqref{e:GPD1} are dressed with light-like Wilson lines to make the quantity gauge invariant; these Wilson lines are equal to unity in the $A^+ = 0$ light cone gauge\footnote{Note that a general GPD depends on the skewness $\xi \sim \Delta^+ / p^+$ with $H^g\equiv H^g(x,\xi, -\Delta_T^2)$ but in these kinematics the skewness is small and is set to zero in our discussion.}.

Altogether, combining this with Eqs.~(\ref{e:dipfact}) and (\ref{e:proton-cs}), one obtains for the cross-section for elastic vector meson production on the proton\footnote{Here and in the following, we ignore saturation corrections that become important when $r_\perp^2 Q_s^2 \sim 1$, where $Q_s$ is the saturation scale in the proton that grows with decreasing $x$. Our formalism can be extended to include such corrections.}
\begin{align} \label{e:CS1}
  \frac{d\sigma^N}{dT} = \frac{1}{4\pi} \left| \int \frac{d^2 r \, d z}{4\pi z (1-z)} 
    \left[ \psi^\gamma \psi^{V *} \right] \!\! (\ul{r}, z) \: \frac{\alpha_s \pi^2}{2} r_T^2 \: 
	H^g_{(N)} (x_{eff} , 0, T) \right|^2\,,
\end{align}
where the averaging over the angles of $\ul{\Delta}$ has now become trivial and we have defined the wave function overlap $[ \psi^\gamma \psi^{V *} ] (\ul{r}, z) \equiv \left[ \sum_{\sigma \sigma'} \psi_{\lambda \sigma \sigma'}^\gamma (\ul{r},z) \left( \psi_{\lambda' \sigma \sigma'}^V (\ul{r}, z) \right)^* \right]$ for brevity.  We note that in the forward limit $T \rightarrow 0$, the GPD $H^g$ reduces down to the ordinary unintegrated gluon distribution: $H^g (x, 0, 0) = x G (x, 1/r_T^2)$, where the inverse dipole size $1/r_T$ provides the transverse momentum cutoff (factorization scale)~\cite{Brodsky:1994kf, Frankfurt:1995jw, Frankfurt:1997fj}.

\subsection{Back-to-Back Electro-disintegration of the Deuteron}
\label{sec:IntroDeut}


\begin{figure}[bt]
 \centering
 \includegraphics[width=\textwidth]{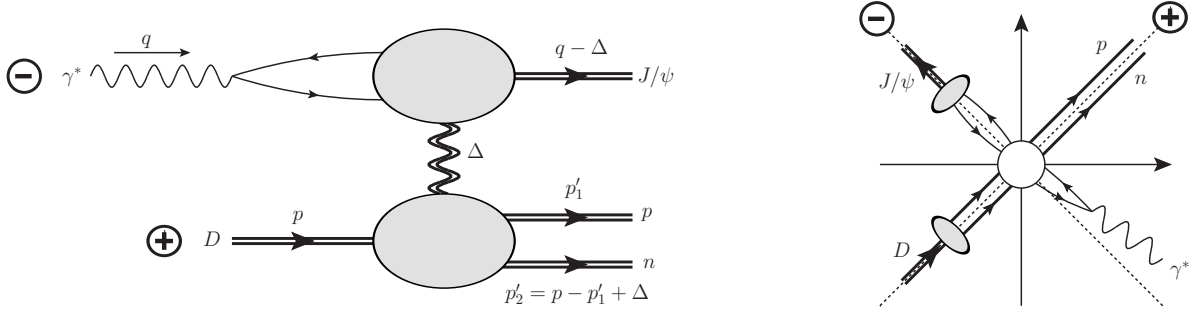}
 \caption{Schematic illustration of the amplitude for diffractive $\Jpsi$ electroproduction on the deuteron: $\gamma_{(q)} + D_{(p)} \rightarrow 
               \Jpsi_{(q-\Delta)} + p _{(p_1^\prime)} + n_ {(p_2^\prime)}$.  (Left Panel):  In Regge kinematics, the photon preferentially fluctuates into a quark 
               dipole and scatters from the deuteron by pomeron exchange (double wavy line in the $t$-channel).  (Right Panel):  Spacetime picture of the
               collision.  The symbols $\oplus$ and $\ominus$ represent the trajectories of the deuteron and photon along the respective light front axes.}
 \label{f:Double_Diffraction} 
\end{figure}

We will now consider diffractive $\Jpsi$ electro- or photo-production on the \textit{deuteron} and require in addition that the deuteron disintegrates into the proton and neutron in the final state with high transverse momentum: $\gamma + D \rightarrow \Jpsi + p + n$.  The scattering is illustrated in Fig.~(\ref{f:Double_Diffraction}) and the kinematics are similar to the proton case,
\begin{align} \label{e:xeff}
  p^\mu &= \left( p^+ , \tfrac{m_D^2}{s} q^-, \ul{0} \right) \hspace{1cm}
  q^\mu = \left( - x p^+, q^-, \ul{0} \right) \notag \\
  \Delta^+ &\approx  - x_{eff} \, p^+ \hspace{1cm}
  \Delta^- \approx \left( \tfrac{p_{1T}^{\prime 2} + m_N^2}{\alpha' (1-\alpha') s} - \tfrac{m_D^2}{s} \right) q^-\,,
\end{align}
with $m_D$ the deuteron mass, $m_N$ the nucleon mass, and $\alpha' \equiv p_1^{\prime +} / p^+$ the fraction of the incoming deuteron momentum carried by the outgoing proton.  In deriving these expressions we have neglected $\Delta_T^2 \ll p_{1T}^{\prime 2}$ as well as $\Delta_T^2 \ll \max(Q^2 , M_V^2)$.  The invariant momentum transfer is $T = \Delta^2 \approx - \Delta_T^2$ and the center-of-mass energy (squared) of the outgoing $NN$ system is then
\begin{align} \label{sNN}
s_{NN} \approx \frac{p_{1T}^{\prime 2} + m_N^2}{\alpha' (1-\alpha')} - m_D^2.
\end{align}

In the limit when $p_{1T}^{\prime 2} \gg \Delta_T^2$, the proton and the neutron emerge nearly back-to-back: $\ul{p_2^\prime} \approx - \ul{p_1^\prime}$.  As the recoil momentum $p_{1T}^\prime$ becomes large, the $NN$ invariant mass $s_{NN}$ also grows; this invariant mass is delivered through the $T$-channel by the component $\Delta^- \approx \tfrac{s_{NN}}{s} \, q^-$.  The amount of energy which can be delivered through the $T$-channel without disturbing the preceding kinematics is limited by the approximation 
\begin{align}
  T = 2 \Delta^+ \Delta^- - \Delta_T^2 \approx - x_{eff} s_{NN} - \Delta_T^2 \approx - \Delta_T^2 \,,
\end{align}
which, in turn, limits the invariant mass in the $NN$ system to $|T| \ll s_{NN} \ll \frac{|T|}{x_{eff}}$ for the approximations to be valid. Thus one needs very high DIS energies for small sized dipoles (with large $M_V^2$ and/or $Q^2$) to probe large relative momenta between the outgoing proton and neutron. 

Following steps identical to those in Section~\ref{sec:protonkin}, introducing in addition the rapidity $y_1^\prime = d p_1^{\prime +} / p_1^{\prime +}$ of the proton and 
the new momentum transfer variable 
\begin{align} \label{e:tdef}
  t \equiv (p_1^\prime - p)^2 =  (1-\alpha^\prime) \left( m_D^2 - \frac{m_N^2}{\alpha^\prime} \right) - \frac{1}{\alpha^\prime} p_{1T}^{\prime 2}
\end{align}
and with $\frac{d^2 p_1^\prime}{(2\pi)^2}=\alpha' \frac{dt}{4\pi}  \frac{ d\phi_1^\prime}{2\pi}$, one can write down the differential cross-section specifying both the overall momentum transfer $T$ and the final state proton kinematics as
\begin{align}
	\frac{d\sigma^D}{dT \, dt \, d y_1^\prime} = \frac{1}{(4\pi)^3} \frac{\alpha'}{1-\alpha'} \int \frac{d\phi_\Delta}{2\pi} \, 
	\frac{d\phi_1^\prime}{2\pi} \, \left| A(\ul{\Delta}, \ul{p_1^\prime} , \alpha') \right|^2 ,
\end{align}
where again the energy rescaled amplitude is $A = M / 2s$.

As with the proton case in Eq.~(\ref{e:dipfact}), the separation of time scales allows us to factorize the photon and vector meson wave functions from the scattering amplitude of the dipole on the deuteron as 
\begin{align} \label{e:dipfact3}
  A (\ul{\Delta}, \ul{p_1^\prime}, \alpha^\prime ) &= \int \frac{d^2 r \, dz}{4\pi z (1-z)} \,
  \left[ \psi^\gamma \psi^{V *} \right] \!\! (\ul{r}, z) \:  \Tr_C \: A^{q \bar q D} (\ul{r}, \ul{\Delta}; \ul{p_1^\prime}, \alpha^\prime).
\end{align}
The propagation of the high energy dipole through the gluon field of the deuteron is unchanged.  However now the nonperturbative matrix element from which the gluon field is taken is not an expectation value in the proton state; it corresponds instead to the transition from the deuteron to the $NN$ system.  Therefore we can define a novel object $\hat{H}^g$ with the same gluon operators as the GPD $H^g$, but evaluated between the deuteron and $NN$ states; this gluon Transition Generalized Parton Distribution (T-GPD) is defined to be
\begin{align}
  \hat{H}^g (x, 0, T ; t ) \equiv \int \frac{dr^-}{2\pi p^+} \, e^{i x p^+ r^-} \bigg\langle p (p_1^\prime) \, n (p + \Delta - p_1^\prime ) \bigg| 
  F^{+i a} (- \tfrac{r}{2}) F^{+ i a} (+ \tfrac{r}{2}) \bigg| D(p) \bigg\rangle \, .
  \label{e:dipfact4}
\end{align}
The relation of the scattering amplitude to the T-GPD is the same as in the proton case, 
\begin{align}
  \Tr_C \, A^{q \bar q D} (\ul{r}, \ul{\Delta}; \ul{p_1^\prime}, \alpha^\prime) &\approx
  \frac{\alpha_s \pi^2}{2} r_T^2 \, \hat{H}^g_{(D)} (x, 0, T; t) \, .
\label{e:dipfact4.1}
\end{align}
Note that we are utilizing this quantity in the limit $|T| \ll |t|$.  The operator expression for Eq.~(\ref{e:dipfact4}) is only valid in the light cone gauge $A^+=0$; in other gauges, it is dressed by light like Wilson lines to preserve gauge invariance.   Since the only major difference from the proton case is the evaluation of the dipole scattering amplitude between different states, we obtain a similar formula,
\begin{align} \label{e:CS2}
  \frac{d\sigma^D}{dT \, dt \, dy_1^\prime} = \frac{1}{(4\pi)^3} \frac{\alpha^\prime}{1 - \alpha'} 
	\bigg| \int \frac{d^2 r \, d z}{4\pi z (1-z)}
        \left[ \psi^\gamma \psi^{V *} \right] \!\! (\ul{r}, z)
        \frac{\alpha_s \pi^2}{2} r_T^2 \, \hat{H}^g_{(D)} (x, 0, T; t) \bigg|^2\,,
\end{align}
which differs only in the kinematic prefactor and the appearance of the T-GPD $\hat{H}^g_{(D)}$.  

As with exclusive $J/\Psi$ production off the proton, the leading order process consists of the projectile dipole exchanging two gluons with deuteron, delivering a sufficient momentum kick to break apart the deuteron. The nonperturbative gluon matrix element \eqref{e:dipfact4} is similar to \eqref{e:GPD1}, but evaluated between the incoming deuteron state $\ket{D}$ and outgoing proton and neutron nucleon states $\bra{N \otimes N}$. In LFPT,  the ``in'' and ``out'' states are evaluated at asymptotic infinity, and the operators are evaluated at $x^+ = 0^+$:
\begin{align}
		&{}_{out}\bra{p n} [F^{+ i a} F^{+ i a}](0^+) \ket{D}_{in} =
		\bra{p n} \mathcal{U}[+\infty^+, 0^+] \: [F^{+ i a} F^{+ i a}](0^+) \:
		\mathcal{U}[0^+, -\infty^+] \ket{D} 
                \notag \\ & \hspace{2cm}
                =\SumInt_{\mathbb{X}}
		\bra{p n} \: \mathcal{U}[+\infty^+, 0^+] \: [F^{+ i a} F^{+ i a}](0^+) \ket{\mathbb{X}} \:\:
		\bra{\mathbb{X}} \mathcal{U}[0^+, -\infty^+] \: \ket{D} \,,
\end{align}
where we have inserted a complete set of states $\ket{\mathbb{X}}$ in the second line.  This decomposition relies on the gluon field insertions being nearly instantaneous relative to the lifetimes of the intermediate states.  In principle, the deuteron wave function has nonzero overlap with an infinite tower of Fock states ($N N, \Delta \Delta, N^8 N^8, \cdots$). However since the deuteron is so loosely bound, the nucleon-nucleon ground state is the dominant configuration and we will assume in the following that the complete set of states $\ket{\mathbb{X}}$ is saturated by this configuration to give 
\begin{align}
        {}_{out}\bra{p n} [F^{+ i a} F^{+ i a}](0^+) \ket{D}_{in} &=
        \notag \\ & \hspace{-2.5cm}
        \int d\Omega_{NN}
	\bra{p n} \: \mathcal{U}[+\infty^+, 0^+] \: [F^{+ i a} F^{+ i a}](0^+)
	\ket{N N \, (\Omega_{NN})}
	\psi^{D \rightarrow N N} (\Omega_{NN})\, ,
\end{align}
where the deuteron LFWF is
\begin{align}
	\psi^{D \rightarrow N N} = \bra{N N} \: \mathcal{U}[0^+, -\infty^+]  \ket{D}.
\end{align}
Like the photon and $\Jpsi$ wave functions, the intrinsic time scales $\sim 1/m_D$ of the deuteron wave function are much longer than the interaction time scale $\sim 1/\sqrt{s}$ with the projectile, leading to a factorization of the matrix element into the wave function times an on-shell $NN$ amplitude.  One therefore obtains for the gluon T-GPD
\begin{align} \label{e:WFconv1}
\hat{H}^g_{(D)} (x,0,T; t) = 
  \int \frac{d \alpha}{4\pi \alpha (1-\alpha)} \frac{d^2 p_1}{(2\pi)^2} \sum_{\sigma_p \sigma_n} \, 
  \psi^D_{\sigma_D ; \, \sigma_p \sigma_n} (\ul{p_1}, \alpha) \, H^g_{\sigma_p^\prime \sigma_n^
	\prime; \, \sigma_p \sigma_n} (x, 0, T; t) \,,
\end{align}
with
\begin{align}
\label{e:WFconv2}
        &H^g_{\sigma_p^\prime \sigma_n^
	\prime; \, \sigma_p \sigma_n} (x, 0, T; t) = \int \frac{d r^-}{2\pi p^+} \, e^{i x p^+ r^-}
	\notag \\ & \times
        \bra{ p_{\sigma_p^\prime} (p_1^\prime) \, n_{\sigma_n^\prime} (p + \Delta - p_1^\prime) } \mathcal{U}[+\infty^+, 0^+] \:
         F^{+i a}(-\thalf r) F^{+i a}(+\thalf r) \ket{ p_{\sigma_p} (p_1) \, n_{\sigma_n} (p - p_1) }  \, .
\end{align}
The polarizations of the incoming deuteron and outgoing proton and neutron are $\sigma_D$, $\sigma_p^\prime$, and $\sigma_n^\prime$, respectively, and the sum over the spins $\sigma_p , \sigma_n$ of the intermediate nucleons is shown explicitly.  

\begin{figure}[hbt]
 \centering
 \includegraphics[width=\textwidth]{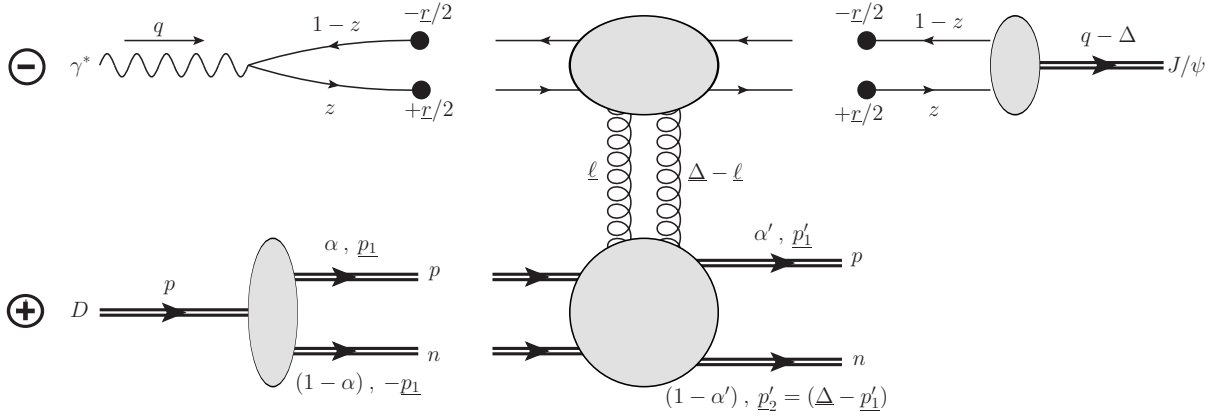}
 \caption{Illustration of the factorization \eqref{e:factoriz} of the cross-section into the wave functions of the virtual photon, vector meson, and deuteron, times the scattering of the dipole on the $NN$ system.}
 \label{f:Factorization}
\end{figure}

The $NN$ matrix element $H^g_{\sigma_p^\prime \sigma_n^\prime ; \, \sigma_p \sigma_n}$ is close to being the standard GPD, except for describing a two-hadron system and therefore retaining dependence on the momentum transfer $t$ to one hadron (the proton) in addition to the momentum transfer $T$ to the center of mass.  In terms of this $NN$ matrix element, the production cross-section can be written
\begin{align}  \label{e:factoriz}
   \frac{d\sigma^D}{dT \, dt \, dy_1^\prime} &= \frac{1}{(4\pi)^3} \frac{\alpha'}{1-\alpha'} \, \Bigg| 
   \int \frac{d^2 r \, d z}{4\pi z (1 - z)}  \left[ \psi^\gamma \psi^{V *} \right] \!\! (\ul{r}, z) \: \frac{\alpha_s \pi^2}{2} r_T^2
   \notag \\ &\times
   \int \frac{d\alpha}{4\pi \alpha (1-\alpha)} \frac{d^2 p_1}{(2\pi)^2}
   \sum_{\sigma_p \sigma_n} \psi^D_{\sigma_D ; \, \sigma_p \sigma_n} (\ul{p_1}, \alpha) \, 
	 H^g_{\sigma_p^\prime \sigma_n^\prime ; \, \sigma_p \sigma_n} (x, 0, T; t) \Bigg|^2,
\end{align}
and is depicted in Fig.~\ref{f:Factorization}.

By considering exclusive vector meson production from DIS on a composite state such as the deuteron, we have introduced another kinematic parameter $t$ into the ``GPD'' $H^g (x,0,T;t)$.  If $t$ is a soft scale, $p_{1T}^{\prime 2} \sim \ord{\Lambda_{QCD}^2}$, then this is simply a nonperturbative distribution.  Because the deuteron is a loose bound state of nucleons, to leading order, the singlet two gluon exchange can be thought of as taking place on one of the nucleons, while the other nucleon is a spectator\footnote{The octet configuration, which we shall soon discuss, where one gluon is exchanged with a proton and the other with the neutron does not apply here since the necessary color neutralizing octet gluon exchange between the two nucleons is forbidden by confinement at large distances.}.  This is equivalent to writing the gluon GPD of the deuteron as the sum of GPD's in each of its nucleons: $\hat{H}^g_{(D)} = 2 H^g_{(N)}$.  In this mechanism, the nucleons are ejected with $p_T^2$ comparable to the momentum transfer $|T|$ from the projectile dipole, or to the intrinsic momentum scale $\sim \Lambda_{QCD}^2$ which the deuteron wave
function can accommodate.

\begin{figure}[bt]
 \centering
 \includegraphics[width=0.7\textwidth]{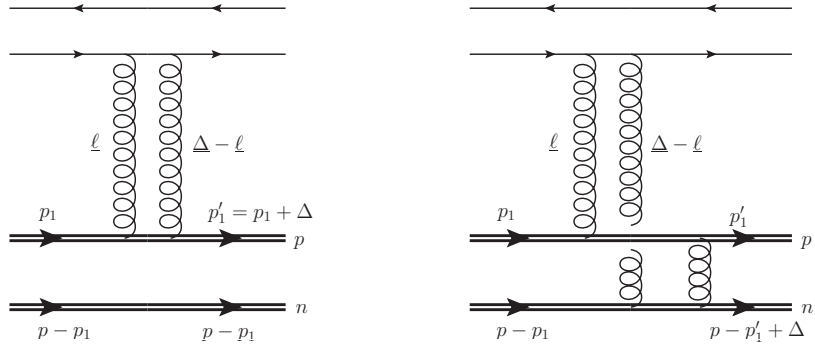}
 \caption{Examples of the dipole-$NN$ $T$-matrix which can contribute to the $|t| \sim |T|$ (left panel) and $|t_1| \gg |T|$ (right panel) regimes.  In the 
               first case, $|t| \approx |T| \approx \Delta_T^2$, and in the second case $|t| \approx p_{1T}^{\prime 2} \gg \Delta_T^2 \approx |T|$.}
 \label{f:LOvsB2B}
\end{figure}

But if the additional kinematic parameter $t$ becomes a {\it hard} scale $|t| \sim p_{1T}^{\prime 2} \gg \Lambda_{QCD}^2$, it must be generated by a short-distance QCD interaction.  Inserting another complete set of states in \eqref{e:WFconv2} (and assuming it is also saturated by the $N N$ state), one can then expand the ``time'' evolution operator \eqref{e:timeev} to lowest nonvanishing order, obtaining 
\begin{align} \label{e:fulldecomp}
  \hat{H}^g_{(D)} (x,0,T; t) &= \int \frac{d \alpha}{4\pi \alpha (1-\alpha)} \frac{d^2 p_1}{(2\pi)^2} 
  \sum_{\sigma_p \sigma_n} \, \psi^D_{\sigma_D ; \, \sigma_p \sigma_n} (\ul{p_1}, \alpha) \: 
  \notag \\ &\times
  \bigg[ \int \frac{d r^-}{2\pi p^+} \, e^{i x p^+ r^-} \int d\Omega_{NN} \:
  \bra{ p_{\sigma_p^\prime} (p_1^\prime) \, n_{\sigma_n^\prime} (p + \Delta - p_1^\prime) }
  V_{NN}^- \ket{ N N } 	
  \notag \\ &\times 
  \frac{1}{\Delta E^-} \times \bra{ N N }| F^{+i a}(-\tfrac{r}{2}) F^{+i a}(+\tfrac{r}{2}) 
  \ket{ p_{\sigma_p} (p_1) \, n_{\sigma_n} (p - p_1) } \bigg] ,
\end{align}
where $V_{NN}^-$ is the first operator in the expansion of $\mathcal{U}$ with a nonvanishing matrix element. Further, $\Delta E^-$ is the energy denominator of the virtual $\ket{NN}$ state between the two-gluon exchange and the $NN$ rescattering.  In this way, we can hope to extract a nucleon-nucleon scattering matrix element from a measurement of the $\Jpsi$ production cross section.

Specifically, the lowest order process consists of the exchange of an additional gluon between the proton and neutron, as shown in the right panel of Fig.~\ref{f:LOvsB2B}.  Normally the exchange of a single gluon does not contribute to $NN$ scattering, since the gluon carries an octet color charge.  However as part of the dipole scattering amplitude, such a process {\it is} possible if each nucleon absorbed exactly one of the gluons from the diffractive $T$-channel exchange with the dipole.  The additional gluon exchange between the nucleons can then neutralize the net octet color charge state each nucleon has acquired from the diffractive pomeron.  

This novel perturbative mechanism, in which the nucleons are temporarily excited into a color octet state, may provide a new window into nucleon-nucleon interactions at short distances.  We will explore this mechanism quantitatively in Sec.~\ref{sec:pQCD}.  While a full understanding of how this perturbative picture 
requires that one demonstrate factorization of the nonperturbative physics beyond leading order,  we will argue that at high $t$ this picture is plausible  because the separation of time scales between the diffractive exchange $\Delta x^+ / p^+_{tot} \sim \ord{1/s}$ and the $NN$ rescattering: $\Delta x^+ / p^+_{NN} \sim \ord{1/p_{1T}^{\prime 2}}$ suggests that a factorization of the T-GPD should survive in a complete treatment.  Regardless, exclusive vector meson production with deuteron breakup into high-$p_T$ nucleons can be used to probe the short distance behavior of the T-GPD $\hat{H}(x,0,T;t)$. The magnitude and $t$ dependence of this object will provide novel information that models of the short range nucleon-nucleon interaction should satisfy. As an example, the multi-Pomeron exchange model~\cite{Yamamoto:2013ada}, currently applied to model neutron star equations of state, should be strongly constrained by this T-GPD.

\section{Perturbative Computation} \label{sec:pQCD}
%
\vspace{-0.1cm}
We will now perform a perturbative computation of the dipole scattering amplitude  $\Tr_C A^{q \bar q D}$ for exclusive vector meson production with deuteron breakup.  In Section~\ref{sec:maincalc} we will simplify the computation greatly by treating each of the nucleons in the scattering as valence quarks. The purpose of this toy model computation is to obtain a feel for the relative importance of initial state versus final state gluon mediated color exchanges, to understand the structure of so-called pinch singularities that can modify naive power counting and to fix numerical coefficients that will be important for our later estimates. 

In Section~\ref{sec:qcount} discuss the contours of the full computation which is challenging even in high-energy asymptotics and and at leading order. While clearly outside the scope of this work, we outline how this computation relates to previous work on large angle high energy elastic scattering. This correspondence will provide an important basis for future quantitative studies. Our study motivates an ansatz for the photo-disintegration cross-section as a product of the modulus squared deuteron wavefunction with the exclusive vector meson photoproduction cross-section and the neutron-proton elastic scattering cross-section. Since each of these quantities can be fixed, they provide a plausible estimate of rates for this process at an EIC.  These will be discussed in Section~\ref{sec:rates}.  

\subsection{Toy model computation} \label{sec:maincalc}
%
\begin{figure}[bt]
 \centering
 \includegraphics[width=0.9\textwidth]{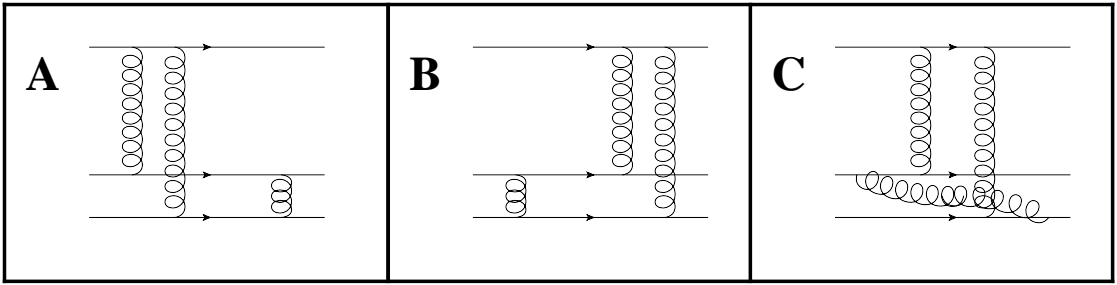}
 \caption{Representative diagrams from the three classes of nucleon-nucleon interactions.  The $NN$ interaction can occur as a final state interaction (A), an initial state interaction (B), or a long lived fluctuation (C).  For each class of diagrams there are also minor variations on the topology, such as altering the order in which the projectile scatters on the two nucleons.}
\label{f:Deut_Table}
\end{figure}
%

It is convenient to perform this calculation using covariant Feynman perturbation theory rather than LFPT, since the on-shell dipole scattering amplitude can be computed in either formalism.  It is also convenient to focus on the diagrams in which both exchanged gluons couple to the quark in the projectile dipole; the generalization to the diagrams involving the antiquark is straightforward.  We will work in Feynman gauge satisfying $\partial_\mu A^\mu = 0$.  

The three general diagrammatic topologies relating to the ordering of the $NN$ interaction with respect to the diffractive exchange with the projectile are illustrated in Fig.~\ref{f:Deut_Table}, by taking the $NN$ interaction to be the exchange of a single gluon at leading order.  In our toy computation, the valence quark  ``nucleons''  are color singlet; this therefore requires that the diffractive exchange with the projectile deliver one gluon to each ``nucleon". The generalization to the scattering amplitude of the full projectile dipole $\Tr_C A^{q \bar q N N}$ will be straightforward. Once this is accomplished, we will convolute the scattering amplitude with the deuteron wave function (as in Eq.~\eqref{e:WFconv1}) to obtain the full dipole scattering amplitude $\Tr_C A^{q \bar q D}$ on the deuteron target.

\begin{figure}[bt]
 \centering
 \includegraphics[width=0.7\textwidth]{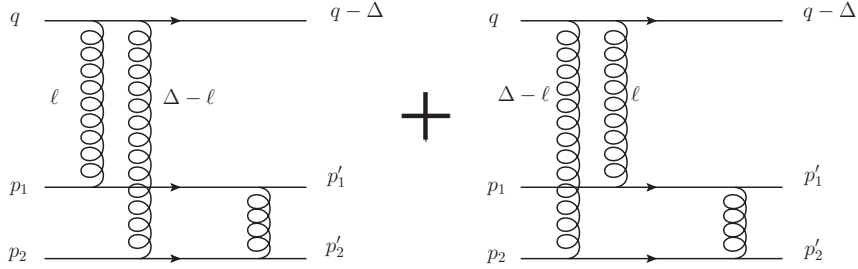}
 \caption{Final state interaction diagrams.  The initial neutron momentum is $p_2 \equiv p - p_1$, and the final neutron momentum is $p_2^\prime = p - p_1^\prime + \Delta$.}
\label{f:FSI_1}
\end{figure}

\begin{figure}[bt]
 \centering
 \includegraphics[width=0.9\textwidth]{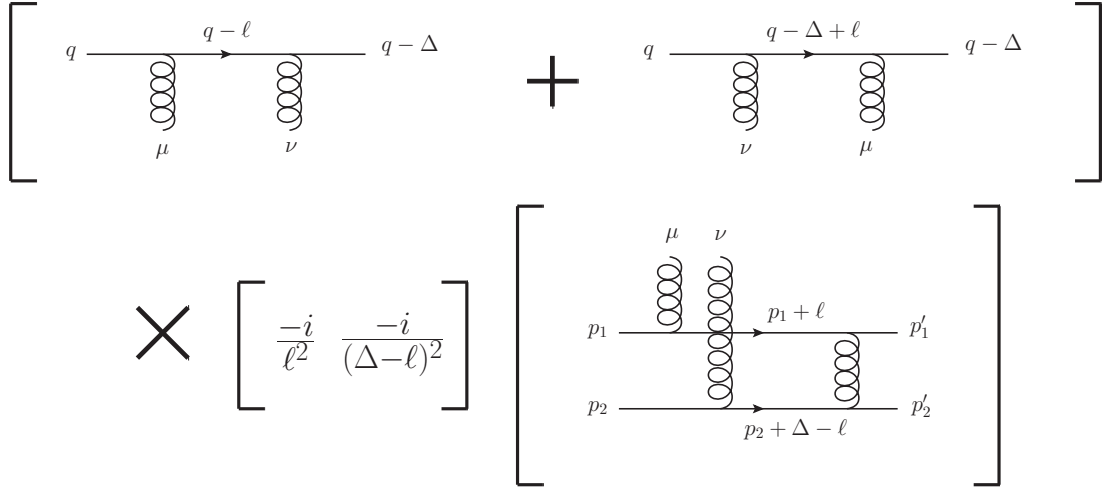}
 \caption{Partition of the diagrams of Fig.~\ref{f:FSI_1} into the upper part $\mathcal{U}_{\mu \nu} (\ell)$ (first bracketed sum of diagrams), lower part $\mathcal{L}^{\mu \nu} (\ell)$ (third bracketed diagram), and gluon propagators.}
\label{f:FSI_2}
\end{figure}

Let us begin by calculating in detail the final state interaction diagrams represented in Fig.~\ref{f:Deut_Table} by category A.  There are two diagrams with this topology, corresponding to the projectile striking the two nucleons in either order; these diagrams are shown in Fig.~\ref{f:FSI_1}.  
%
%
By labeling the momenta and indices appropriately, it is possible to arrange the two diagrams so that they only differ in the flow of momentum through the projectile quark; this conveniently allows us to write the sum of the diagrams as
\begin{align}
	i \, \Tr_C A^{q N N} \, \delta_{\sigma \sigma'} &= \frac{1}{2 s}  \frac{C_F}{4 N_c}\int \frac{d^4 \ell}{(2\pi)^4} \mathcal{U}_{\mu \nu} (\ell) \left( \frac{-i}{\ell^2} \frac{-i}{(\Delta - \ell)^2} \right) \mathcal{L}^{\mu \nu} (\ell)
\end{align}
with $\mathcal{U}_{\mu \nu}$ the upper part and $\mathcal{L}^{\mu \nu}$ the lower part of the diagrams, as illustrated in Fig.~\ref{f:FSI_2}.  
%

For the upper part of the diagram, we sum the two pieces shown in Fig.~\ref{f:FSI_2} and obtain
\begin{align}
	\mathcal{U}_{\mu \nu} (\ell) &\equiv \ubar{\sigma'} (q-\Delta) [i g \gamma_\nu]
	\left[\frac{i (\slashed{q}-\slashed{\ell} + m_q)}{(q-\ell)^2 - m_q^2 + i\epsilon} \right]
	[i g \gamma_\mu] U_\sigma (q)
	\notag \\ &+
	\ubar{\sigma'} (q-\Delta) [i g \gamma_\mu]
	\left[\frac{i (\slashed{q}-\slashed{\Delta}+\slashed{\ell} + m_q)}{(q-\Delta+\ell)^2 - m_q^2+ i \epsilon} \right] [i g \gamma_\nu] U_\sigma (q),
\end{align}
where $\sigma (\sigma')$ is the initial (final) spin of the projectile quark and $m_q$ is its mass. To simplify the expression, we take the eikonal part $(q^- \gamma^+)$ of the projectile quark propagator in the numerator, assuming that $|\ell^-| \ll q^-$:
\begin{align}
\mathcal{U}_{\mu \nu} (\ell) &= 
	-i g^2 q^- \left\{
	\frac{\ubar{\sigma'}(q-\Delta) \, \gamma_\nu \gamma^+ \gamma_\mu U_\sigma (q)}{(q-\ell)^2 - m_q^2 + i\epsilon} +
	\frac{\ubar{\sigma'}(q-\Delta) \, \gamma_\mu \gamma^+ \gamma_\nu U_\sigma (q)}{(q-\Delta+\ell)^2 - m_q^2 + i\epsilon}
	\right\}.
\end{align}
The dominant contribution to the numerator algebra comes from the gamma matrix structure $\gamma_\nu \gamma^+ \gamma_\mu \rightarrow  \delta_\mu^+ \delta_\nu^+ \, \gamma^- \gamma^+ \gamma^- = 2 \delta_\mu^+ \delta_\nu^+  \, \gamma^-$, where we have picked out the eikonal quark/gluon vertices and used the anticommutation relations.  Using this brings the upper part of the diagram into the form
\begin{align}
	\mathcal{U}_{\mu \nu} (\ell) &= 
	-2 i g^2 q^- \delta_\mu^+ \delta_\nu^+ \, [ \ubar{\sigma'}(q-\Delta) \gamma^- U_\sigma (q)] 
        \notag \\ & \hspace{1cm} \times
	\left\{	\frac{1}{(q-\ell)^2 - m_q^2 + i\epsilon} + \frac{1}{(q-\Delta+\ell)^2 - m_q^2 + i\epsilon}
	\right\}.
\end{align}
The leading part of the spinor matrix element is $\ubar{\sigma'} (q-\Delta) \gamma^- U_\sigma (q) \approx 2 q^- \delta_{\sigma \sigma'}$, which simplifies the expression to
\begin{align}
	\mathcal{U}_{\mu \nu} (\ell) &\approx
	-2 i g^2  q^- \delta_\mu^+ \delta_\nu^+ \delta_{\sigma \sigma'} \, 
	\left\{	- \frac{1}{\ell^+ - \ell_1^+ - i\epsilon} + \frac{1}{\ell^+ - \ell_2^+ +i\epsilon}
	\right\},
\end{align}
where the poles of the propagators are given by
\begin{align}
	\ell_1^+ = q^+ - \frac{(q-\ell)_T^2 + m_q^2}{2 q^-} \hspace{1cm}
	\ell_2^+ = -(q-\Delta)^+ + \frac{(q-\Delta - \ell)_T^2 + m_q^2}{2 q^-} ,
\end{align}
with the projectile quarks treated as (nearly) on-shell.  (Recall that this was a consequence of the separation of time scales.)  Using this, we rewrite the amplitude as 
\begin{align} \label{e:Low0}
	\Tr_C A^{q N N} = \frac{g^2 C_F}{8 N_c} \frac{1}{p^+} 
	\int\frac{d^4\ell}{(2\pi)^4} \frac{1}{\ell_T^2 (\Delta-\ell)_T^2} 
	\left[\frac{1}{\ell^+ - \ell_2^+ + i\epsilon} -
	\frac{1}{\ell^+ - \ell_1^+ - i\epsilon} \right] \mathcal{L}^{++}(\ell),
\end{align}
where we have taken the $t$-channel gluons to be Glauber / Coulomb gluons $(|\ell^+ \ell^-| \ll \ell_T^2)$.  Since the upper part of the diagram involving the projectile dipole is the same among all the diagrams in Fig.~\ref{f:Deut_Table}, we can use \eqref{e:Low0} as the starting point for the calculation of all of them.

The lower part of the final state interaction diagram in Fig.~\ref{f:FSI_2}, taken in the eikonal limit $\delta_\mu^+ \delta_\nu^+ \mathcal{L}^{\mu\nu} (\ell) = \mathcal{L}^{++} (\ell)$, is given by
\begin{align}
	&\mathcal{L}_A^{++}(\ell) \equiv \ubar{\sigma_p^\prime} (p_1^\prime) [i g \gamma_\alpha]
	\left[\frac{i (\slashed{p_1} + \slashed{\ell} + m_N)}{(p_1 + \ell)^2 - m_N^2 + i \epsilon}\right] 
	[i g \gamma^+] U_{\sigma_p} (p_1)
	\ubar{\sigma_n^\prime} (p_2^\prime) [i g \gamma^\alpha]\notag \\
	&\times \left[\frac{i (\slashed{p_2} + \slashed{\Delta} - \slashed{\ell} + m_N)}{(p_2 + \Delta - \ell)^2 - m_N^2 + i \epsilon}\right] [i g \gamma^+] U_{\sigma_n} (p_2)
	\left[\frac{-i}{(p_1 + \ell - p_1^\prime)^2 + i\epsilon} \right] \,,
\end{align}
where $\sigma_p (\sigma_n)$ are the initial proton (neutron) spins and $\sigma_p^\prime (\sigma_n^\prime)$ are the final spins, and we use the subscript $A$ to indicate the category from Fig.~\ref{f:Deut_Table}.  Again keeping the eikonal part of the quark propagators $p_{1 (2)}^+ \gamma^-$ for $|\ell^+| \ll p_{1(2)}^+$, we reduce the expression to
\begin{align} \label{e:Low1}
	\mathcal{L}_A^{++}(\ell) &= \frac{ i g^4 \, p_1^+ p_2^+ }{[(p_1 + \ell)^2 - m_N^2 + i \epsilon][(p_2 + \Delta - \ell)^2 - m_N^2 + i \epsilon][(p_1 + \ell - p_1^\prime)^2 + i\epsilon]}
	\notag \\ &\times
	[\ubar{\sigma_p^\prime} (p_1^\prime) \gamma_\alpha \gamma^- \gamma^+ U_{\sigma_p} (p_1)] [\ubar{\sigma_n^\prime} (p_2^\prime) \gamma^\alpha \gamma^- \gamma^+ U_{\sigma_n} (p_2)].
\end{align}
In this eikonal approximation for the quark propagators, the only nonzero contribution from the final state rescattering is from $\gamma_\alpha = \gamma_\bot$.  The numerator structure reduces to
\begin{align}
	\mathrm{Num} &\equiv [\ubar{\sigma_p^\prime} (p_1^\prime) \gamma_\alpha \gamma^- \gamma^+ U_{\sigma_p} (p_1)] \, [\ubar{\sigma_n^\prime} (p_2^\prime) \gamma^\alpha \gamma^- \gamma^+ U_{\sigma_n} (p_2)]
	\notag \\&=
	-[\ubar{\sigma_p^\prime} (p_1^\prime) \gamma_\bot^i \gamma^- \gamma^+ U_{\sigma_p} (p_1)] \, [\ubar{\sigma_n^\prime} (p_2^\prime) \gamma_\bot^i \gamma^- \gamma^+ U_{\sigma_n} (p_2)],
\end{align} 
and a direct evaluation of the matrix element yields
\begin{align} \label{e:Spinors1}
	\ubar{\sigma_p^\prime} (p_1^\prime) \gamma_\bot^i \gamma^- \gamma^+ U_{\sigma_p} (p_1) &= 
        2 \sqrt{\frac{p_1^+}{p_1^{\prime +}}} \Big\{ (\delta^{i j}-i\sigma_p \epsilon_T^{i j}) p_{1 \bot}^{\prime j} \delta_{\sigma_p \sigma_p^\prime}
        - i m_N \epsilon_T^{i j} [\sigma_\bot^j]_{\sigma_p^\prime \sigma_p} \Big\}
\end{align}
using the spinors of \cite{Lepage:1980fj} ($[\vec\sigma]$ are the Pauli matrices, and $\epsilon_T^{ij}$ is the two-dimensional Levi-Civita tensor).  

We can simplify this expression by restricting the range of longitudinal momenta of the ``nucleons''.  The deuteron wave function is largely nonrelativistic, with the longitudinal momentum shared roughly equally among the nucleons: $p_1^+ \approx p_2^+ \approx \half p^+$.  If we consider the produced nucleon in the same vicinity $p_1^{\prime +} \approx \half p^+$, then the longitudinal momentum of the nucleons is roughly unchanged: $p_1^+ \approx p_1^{\prime +}$.  

Since we are interested in a hard transverse momentum in the final state
\footnote{In principle, a hard transverse momentum can also exist in the initial state such that 
$p_{1T}^{\prime 2}\sim p_{1T}^2 \gg m_N^2$.  This corresponds to the high-$p_T$ tail of the deuteron wave function, which is mimicked in this model by the initial-state interaction diagrams (class B of Fig.~\ref{f:Deut_Table}).} 
we will take $p_{1T}^{\prime 2} \gg p_{1T}^2, m_N^2$, which gives 
\begin{align} \label{e:Spinors2}
	&\ubar{\sigma_p^\prime} (p_1^\prime) \gamma_\bot^i \gamma^- \gamma^+ U_{\sigma_p} (p_1) = 
        2  (\delta^{i j}-i\sigma_p \epsilon_T^{i j}) p_{1 \bot}^{\prime j} \delta_{\sigma_p \sigma_p^\prime}  \notag \\
	&\ubar{\sigma_n^\prime} (p_2^\prime) \gamma_\bot^i \gamma^- \gamma^+ U_{\sigma_n} (p_2) =
	- 2  (\delta^{i k}-i\sigma_n \epsilon_T^{i k}) p_{1 \bot}^{\prime k} \delta_{\sigma_n \sigma_n^\prime} ,,
\end{align}
where we have used $\ul{p_2^\prime} = \ul{p} - \ul{p_1^\prime} + \ul{\Delta} \approx - \ul{p_1^\prime}$.  Multiplying the two spinor matrix elements, we obtain for the numerator factor
\begin{align} \label{e:Spinors2.5}
	\mathrm{Num} &= 
	-[\ubar{\sigma_p^\prime} (p_1^\prime) \gamma_\bot^i \gamma^- \gamma^+ U_{\sigma_p} (p_1)] \, [\ubar{\sigma_n^\prime} (p_2^\prime) \gamma_\bot^i \gamma^- \gamma^+ U_{\sigma_n} (p_2)]	
	\notag \\ &=
	+8 p_{1 T}^{\prime 2} \: \delta_{\sigma_p^\prime \sigma_p} \delta_{\sigma_n^\prime \sigma_n} 
	\delta_{\sigma_p , - \sigma_n}\, .
\end{align}
The resulting spin structure is interesting; in addition to the helicity preserving eikonal scattering which keeps $\sigma_{p (n)}^\prime = \sigma_{p (n)}$, the expression couples to the component $\sigma_p = - \sigma_n$ of the deuteron wave function.  Inserting this back into Eq.~\eqref{e:Low1}, we obtain
\begin{align}
	\mathcal{L}_A^{++} (\ell) = \frac{+8 i g^4 \, p_1^+ p_2^+ \, p_{1T}^{\prime 2} \:\: \delta_{\sigma_p^\prime \sigma_p} \delta_{\sigma_n^\prime \sigma_n} 
	\delta_{\sigma_p , - \sigma_n}}{[(p_1 + \ell)^2 -m_N^2 + i \epsilon][(p_2 + \Delta - \ell)^2 -m_N^2 + i \epsilon][(p_1 + \ell - p_1^\prime)^2 + i\epsilon]}.
\end{align}
The final state interaction gluon $(p_1 + \ell - p_1^\prime)^\mu$ has longitudinal $+ , -$ components which are both small--it is a Glauber gluon. We therefore keep only the transverse momentum $(p_1 + \ell - p_1^\prime)_T^2 \approx - p_{1 T}^{\prime 2}$ and neglect the pole in the longitudinal integrals.  This simplifies the expression down to
\begin{align} \label{e:Low2}
	\mathcal{L}_A^{++} (\ell) 
= \frac{+ 2 i g^4 \:\: \delta_{\sigma_p^\prime \sigma_p} \delta_{\sigma_n^\prime \sigma_n} \delta_{\sigma_p , - \sigma_n}}{[\ell^- - \ell_3^- + i\epsilon][\ell^- - \ell_4^- - i\epsilon]}\, ,
\end{align}
where
\begin{align} \label{e:pnpoles1}
	\ell_3^- = - p_1^- +  \frac{(p_1 + \ell)_T^2 + m_N^2}{2 p_1^+} \hspace{1cm}
	\ell_4^- &= p_2^- + \Delta^- - \frac{(p_2 + \Delta - \ell)_T^2 + m_N^2}{2 p_2^+}
\end{align}
are the poles of the ``proton" and ``neutron" propagators respectively.  

\begin{figure}[bt]
 \centering
 \includegraphics[width=0.9\textwidth]{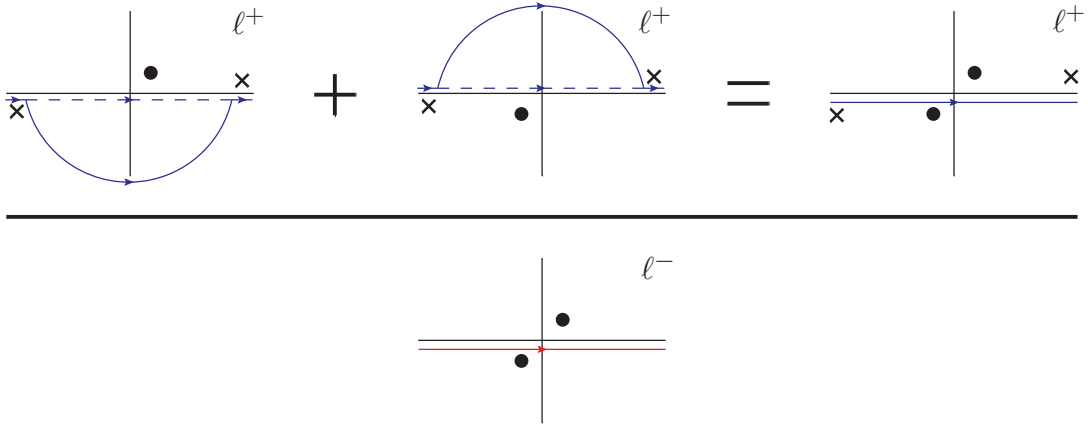}
 \caption{Pole structure in the complex $\ell^\pm$ plane.  The $\ell^+$ pole (upper panel) corresponds to the intermediate quark propagator from the projectile dipole, with a contour that can be deformed far away from the origin.  The magnitude of the deformation is limited by the presence of auxiliary poles (denoted by crosses) from the $t$-channel gluons.  The $\ell^-$ poles (lower panel) correspond to the intermediate proton/neutron propagators.}
\label{f:Poles_1}
\end{figure}

Combining Eqs.~\eqref{e:Low0} and \eqref{e:Low2}, we obtain
\begin{align} \label{e:Low3}
  &\Tr_C A^{q N N}_{(A)} = \frac{g^6 C_F}{4 N_c} \frac{1}{p^+} \:\: (\delta_{\sigma_p^\prime \sigma_p} \delta_{\sigma_n^\prime \sigma_n} \delta_{\sigma_p , - \sigma_n})
	\notag \\ & \hspace{0.75cm}
	\times \int\frac{d^4 \ell}{(2\pi)^4} \frac{i}{\ell_T^2 (\Delta - \ell)_T^2} 
	\left[\frac{1}{\ell^+ - \ell_2^+ + i\epsilon} - \frac{1}{\ell^+ - \ell_1^+ - i\epsilon} \right] \frac{1}{\ell^- - \ell_3^- + i\epsilon} \frac{1}{\ell^- - \ell_4^- - i\epsilon}.
\end{align}
The pole structure which remains in the integral is important; it determines not only the value of the expression, but also the virtuality - and hence the lifetime - of the propagators (see Fig.~\ref{f:Poles_1}).  

The arguments below follow the discussion of factorization and pinched poles in \cite{Collins:2011zzd}.
%
%
The poles $\ell_3^-$ and $\ell_4^-$ of $\ell^-$ generated by our proton and neutron propagators are ``pinched'' (Fig.~\ref{f:Poles_1}, lower panel): they occur at parametrically small values $\ell_{3(4)}^- \propto 1 / p^+$ and lie on opposite sides of the real axis.  This means that the integration contour along the real $\ell^-$ axis cannot be deformed to avoid the poles at $\ell^- \approx 0$, and, correspondingly, the proton / neutron propagators are trapped to have parametrically small virtualities $(p_1 + \ell)^2 , (p_2 + \Delta - \ell)^2 \sim p_{1T}^2 , p_{1T}^{\prime 2}$.  When expressed as a dimensionless ratio compared to the hardest scales in the problem, e.g. $(p_1 + \ell)^2 / s$ or $(p_1 + \ell)^2 / \langle k_T^2 \rangle$ with $ \langle k_T^2 \rangle \sim z (1-z) Q^2 + m_q^2$, the relative virtuality of these propagators goes to zero.  So it seems that, similar to the case of collinear factorization, the proton/neutron propagators can be well approximated as being on-shell, with long lifetimes proportional to the inverse of the virtuality.

\begin{figure}[bt]
 \centering
 \includegraphics[width=0.5\textwidth]{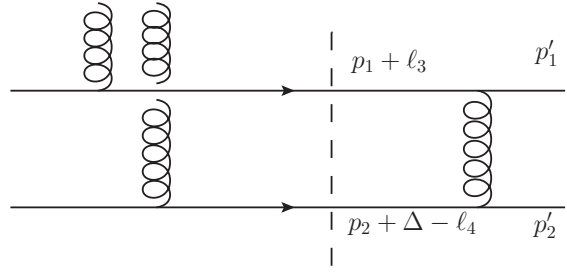}
 \caption{Energy denominator of the intermediate $p n$ state (dashed line) for the final-state interaction topology $\Delta E_{FSI}$ representing the ``residual off-shell-ness'' of the intermediate state.}
\label{f:FSI_cut}
\end{figure}

A more unusual case is given by the poles $\ell_1^+$ and $\ell_2^+$ of $\ell^+$ generated by the propagator of the projectile quark in the two diagrams shown in Fig.~\ref{f:FSI_1}.  In this case, the two poles correspond to two different diagrams, associated with the flow of $\ell^+$ in either direction through the projectile quark propagator.  Because the two poles are not simultaneously imposed in the same diagram, it is possible to deform the $\ell^+$ contour of integration away from the real axis such that $\ell^+$ is not parametrically small anywhere on the contour (Fig.~\ref{f:Poles_1}, upper panel). However, despite the fact that the $\ell^+$ integration contours are not trapped in the individual diagrams, the result of adding the two diagrams together in this fashion is to effectively trap the contour between the two poles (Fig.~\ref{f:Poles_1}, upper panel).  By explicitly adding the bracketed terms in \eqref{e:Low3}, we obtain
\begin{align} \label{e:Int1}
	\mathcal{I} &\equiv \int\frac{d^4 \ell}{(2\pi)^4} \frac{i}{\ell_T^2 (\Delta - \ell)_T^2} \left[\frac{1}{\ell^+ - \ell_2^+ + i\epsilon} - \frac{1}{\ell^+ - \ell_1^+ - i\epsilon} \right] \frac{1}{\ell^- - \ell_3^- + i\epsilon} \frac{1}{\ell^- - \ell_4^- - i\epsilon}
	\notag \\ &=
	\int\frac{d^4 \ell}{(2\pi)^4} \frac{i}{\ell_T^2 (\Delta - \ell)_T^2} 
	\left[\frac{\ell_2^+ - \ell_1^+}{[\ell^+ - \ell_1^+ - i\epsilon][\ell^+ - \ell_2^+ + i\epsilon]}\right] 
	\frac{1}{\ell^- - \ell_3^- + i\epsilon} \frac{1}{\ell^- - \ell_4^- - i\epsilon}.
\end{align}
Although it is not a physical pinch corresponding to a long-lived intermediate state, the sum of the diagrams generates an effective pinch which can be used to evaluate the integral. One obtains   
\begin{align} \label{e:Low4}
	\mathrm{Im}~\Tr_C A^{q N N}_{(A)} = 
	\frac{g^6 C_F}{4 N_c} \left(\frac{1}{p^+  (\ell_4^- - \ell_3^-)}\right)  \:\: 
	(\delta_{\sigma_p^\prime \sigma_p} \delta_{\sigma_n^\prime \sigma_n} \delta_{\sigma_p , - \sigma_n}) \int \frac{d^2 \ell}{(2\pi)^2} \frac{1}{\ell_T^2 (\Delta - \ell)_T^2},
\end{align}
where in the eikonal approximation used here, the real part of the amplitude is zero.  The factor $1 / p^+  (\ell_4^- - \ell_3^-)$ is the residual effect of the proton/neutron poles, arising from collecting the residue of one pole (putting it on shell) and obtaining the resulting ``off-shell-ness'' of the other pole.  Evaluating this factor, one observes that 
the ``residual off-shell-ness'' of the intermediate $p n$ state is just the difference between the minus momentum of the on-shell final-state $p n$ system $(p_1^{\prime -} + p_2^{\prime -})$ and the minus momentum of the intermediate state, evaluated with both nucleons on shell: $(p_1 + \ell_3)^- + (p_2 + \Delta - \ell_4)^-$. This is exactly  the energy denominator from LFPT of the intermediate state (see Fig.~\ref{f:FSI_cut} and compare with Eq.~\eqref{e:fulldecomp}): $\ell_4^- - \ell_3^- = p_1^{\prime -} + p_2^{\prime -} - (p_1 + \ell_3)^- - (p_2 + \Delta - \ell_4)^- \equiv \Delta E_{FSI}^-$. 
%
If we take $p_{1T}^{\prime 2} \gg p_{1T}^2 , \ell_T^2 , \Delta_T^2$, then the energy denominator is dominated by the minus momentum of the final state $\Delta E_{FSI}^- \approx + E_{final}^-$ and
\begin{align}
  \ell_4^- - \ell_3^- \approx \frac{p_{1T}^{\prime 2}}{2 p_1^+} + \frac{p_{1T}^{\prime 2}}{2 p_2^+}
  \approx \frac{p_{1T}^{\prime 2}}{2 \alpha ( 1-\alpha) p^+},
\end{align}
where $\alpha \equiv p_1^+ / p^+$ is the momentum fraction of the deuteron carried by the proton.  We then obtain
\begin{align} \label{e:Low4.1}
	\!\!\!\!
        \mathrm{Im}~\Tr_C A^{q N N}_{(A)} = 
	+8\pi \frac{\alpha_s^3 C_F}{N_c} \frac{\alpha (1-\alpha)}{p_{1T}^{\prime 2}}
        \:\: (\delta_{\sigma_p^\prime \sigma_p} \delta_{\sigma_n^\prime \sigma_n} \delta_{\sigma_p , - \sigma_n}) \:
	\int d^2 x \, e^{-i \ul{\ell} \cdot \ul{x}} \, e^{-i(\ul{\Delta}-\ul{\ell})\cdot\ul{x}} \ln^2\frac{1}{x_T \Lambda}	\,,
\end{align}
where we have replaced the transverse momentum integral in Eq.~\eqref{e:Low4} by its coordinate space equivalent.  The suggestive form of the Fourier factors in \eqref{e:Low4.1} was chosen to show that both gluons interact with the projectile quark at transverse position $\ul{x}$.  The generalization to the case of the full dipole is then straightforward (compare with Eq.~\eqref{e:protondip1}), yielding for the final state interaction topology (A): 
\begin{align} \label{e:FSI_ans}
	\mathrm{Im} \, \Tr_C A^{q \bar q N N}_{(A)} &\approx
	+8\pi \frac{\alpha_s^3 C_F}{N_c} \frac{\alpha (1-\alpha)}{p_{1T}^{\prime 2}}
        \:\: (\delta_{\sigma_p^\prime \sigma_p} \delta_{\sigma_n^\prime \sigma_n} \delta_{\sigma_p , - \sigma_n}) 
	\int d^2 b \, e^{-i \ul{\Delta} \cdot \ul{b}} \bigg(
	\ln^2\frac{1}{|b - z r|_T \Lambda} 
	\notag \\ &
	- 2 \ln\frac{1}{|b - z r|_T \Lambda} \ln \frac{1}{|b + (1-z)r|_T \Lambda}
	+ \ln^2 \frac{1}{|b + (1-z)r|_T \Lambda} \bigg),
\end{align}
where we have kept the dominant imaginary part of the amplitude.

\begin{figure}[bt]
 \centering
 \includegraphics[width=0.6\textwidth]{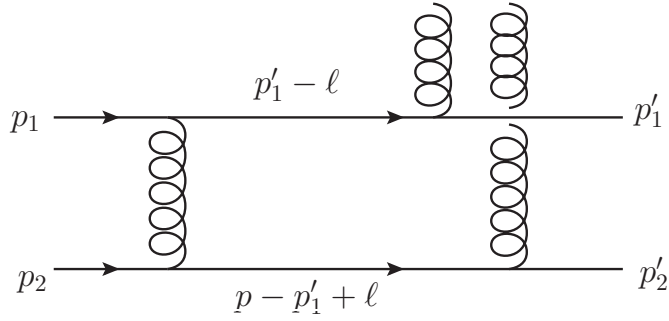}
 \caption{The lower half of the diagram $\mathcal{L}^{++}$ for the initial-state interaction topology (Panel B of Fig.~\ref{f:Deut_Table}).}
\label{f:ISI_1}
\end{figure}

Now let us consider the initial state interaction diagrams (Panel B of Fig.~\ref{f:Deut_Table}).  The treatment of the upper part of the diagram $\mathcal{U}_{\mu \nu}$ is the same as before, so we can begin with Eq.~\eqref{e:Low0} and the lower part of the diagram shown in Fig.~\ref{f:ISI_1}.  Eikonalizing the ``valence" proton and neutron propagators, as previously, using the approximation $p_1^{\prime +} \approx p_1^+$ and recognizing $(p_1+\ell-p_1^\prime)^2 \approx - p_{1T}^{\prime 2}$ to be a Glauber gluon propagator, we find 
\begin{align} \label{e:Low4.5}
        \mathcal{L}_B^{++}(\ell) = 
        \frac{-\tfrac{1}{4} i g^4 \:\:
        [\ubar{\sigma_p^\prime} (p_1^\prime) \gamma^+ \gamma^- \gamma_\bot^i U_{\sigma_p} (p_1)] 
        [\ubar{\sigma_n^\prime} (p_2^\prime) \gamma^+ \gamma^- \gamma_\bot^i U_{\sigma_n} (p_2)]
        }
        {[\ell^- - \ell_5^- - i \epsilon] [\ell^- - \ell_6^- + i \epsilon] \: p_{1T}^{\prime 2}} ,
\end{align}
where the poles of the proton and neutron propagators, respectively, are
\begin{align} \label{e:pnpoles2}
	\ell_5^- = p_1^{\prime -} - \frac{(p_1^\prime - \ell)_T^2 + m_N^2}{2 p_1^+} \,\,; \,\,
	\ell_6^- &= p_1^{\prime -} - p^- + \frac{(p-p_1^\prime+\ell)_T^2 + m_N^2}{2 p_2^+}\,.
\end{align}

The form $\bar{U} (p_1^\prime) \gamma^+ \gamma^- \gamma_\bot U(p_1)$ in Eq.~\eqref{e:Low4.5} is different from the form $\bar{U} (p_1^\prime) \gamma_\bot \gamma^- \gamma^+ U(p_1)$ which appeared for the final state interaction topology.  Direct evaluation of the new matrix element yields
\begin{align} \label{e:Spinors4}
	\ubar{\sigma_p^\prime} (p_1^\prime) \gamma^+ \gamma^- \gamma_\bot^i U_{\sigma_p} (p_1) &=
	2 \sqrt{\frac{p_1^{\prime +}}{p_1^+}} \Big\{
        (\delta^{i j} + i \sigma_p \epsilon_T^{i j}) p_{1\bot}^j \delta_{\sigma_p \sigma_p^\prime} + i m_N \epsilon_T^{i j} [ \sigma_\bot^j ]_{\sigma_p^\prime \sigma_p}
        \Big\} .
\end{align}
In the final state interaction topology, the numerator structure $\bar{U} (p_1^\prime) \gamma_\bot \gamma^- \gamma^+ U(p_1)$ coupled to the transverse momentum $p_{1\bot}^\prime$ in the {\it{final state}}, which is {\it{large}}; in the the initial state interaction topology considered here, the numerator structure $\bar{U} (p_1^\prime) \gamma^+ \gamma^- \gamma_\bot U(p_1)$ couples to the transverse momentum $p_{1 \bot}$ in the {\it{initial state}}, which is {\it{small}}.  As a consequence, in the limits $p_1^{\prime +} \approx p_1^+$ and $p_{1T}^{\prime 2} \gg p_{1T}^2 , m_N^2$ being considered, the complete numerator of Eq.~\eqref{e:Low4.5} remains fixed at $\ord{m_N^2}$ instead of becoming large as $\ord{p_{1T}^{\prime 2}}$ as in Eq.~\eqref{e:Spinors2.5}.  Since the gluon propagator $(p_1+\ell-p_1^\prime)^2 \approx - p_{1T}^{\prime 2}$ itself is large, this leads to a suppression by  $\ord{m_N^2 / p_{1T}^{\prime 2}}$ of the initial state interaction diagrams relative to the final state interaction diagrams.

One can check that the integrals over the poles in the initial state interaction topology are comparable to those in the final state interaction expression. Thus since the numerator algebra has introduced a suppression factor of $\ord{m_N^2 / p_{1T}^{\prime 2}}$, we can neglect the contributions from the initial state interaction topology (category B from Fig.~\ref{f:Deut_Table}) when the transverse momentum of the detected proton is much larger than $1$ GeV.

\begin{figure}[bt]
 \centering
 \includegraphics[width=0.9\textwidth]{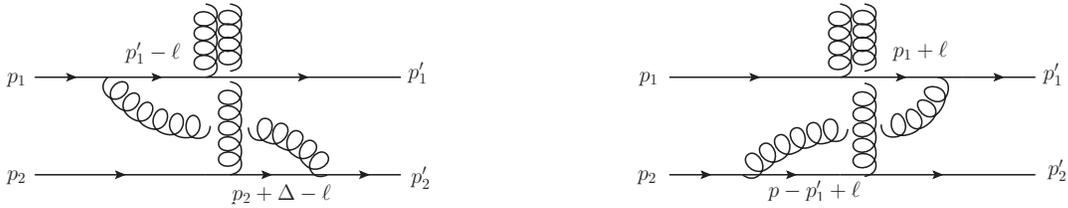}
 \caption{Two diagrams for the ``long-lived fluctuation'' topology (Panel C in Fig.~\ref{f:Deut_Table}).}
\label{f:LLF}
\end{figure}

A similar analysis to the initial and final state cases can be carried out for the ``long lived fluctuation'' diagrams (Panel C in Fig.~\ref{f:Deut_Table}, leading to the lower parts shown in Fig.~\ref{f:LLF}). As for the initial state interaction case, there is a suppression by $\ord{\frac{m_N}{p_{1T}^\prime}}$ coming from the $\bar{U} \gamma_\bot \gamma^- \gamma^+ U$ structure.  Furthermore (and more importantly), the combination of poles which appear in these long lived fluctuations are {\it{both on the same side of the real axis}}.  This breaks the pinch of the $\ell^-$ poles which was present before, allowing us to close the $\ell^-$ integration contour on the opposite side of the real axis and obtain zero for the integral (with eikonal accuracy).  Thus these diagrams in category C are far more suppressed than the initial state interaction diagrams and can therefore be neglected entirely.

As a result of these considerations, we find that the dominant diagrams for $p_{1T}^{\prime 2} \gg m_N^2$ are the final state interaction diagrams (category A from Fig.~\ref{f:Deut_Table}), so that the leading contribution at this order is then given by Eq.~\eqref{e:FSI_ans}:
\begin{align}
  \Tr_C\,A^{q \bar q NN} &=
  +8\pi i \frac{\alpha_s^3 C_F}{N_c} \frac{\alpha (1-\alpha)}{p_{1T}^{\prime 2}}
        \:\: (\delta_{\sigma_p^\prime \sigma_p} \delta_{\sigma_n^\prime \sigma_n} \delta_{\sigma_p , - \sigma_n}) 
	\int d^2 b \, e^{i \ul{\Delta} \cdot \ul{b}} \bigg(
	\ln^2\frac{1}{|b - z r|_T \Lambda} 
	\notag \\ &
	- 2 \ln\frac{1}{|b - z r|_T \Lambda} \ln \frac{1}{|b + (1-z)r|_T \Lambda}
	+ \ln^2 \frac{1}{|b + (1-z)r|_T \Lambda} \bigg).
\end{align}
The dipole scattering amplitude on the deuteron is simply obtained by convoluting this result with the deuteron wave function:
\begin{align}
  \Tr_C\,A^{q \bar q D} = \sum_{\sigma_p \sigma_n} \int\frac{d\alpha}{4\pi \alpha(1-\alpha)} \frac{d^2 p_1}{(2\pi)^2} 
  \psi_{\sigma_D ; \, \sigma_p \sigma_n}^D (\ul{p_1} , \alpha) \times \Tr_C A^{q \bar q NN} \, .
\label{e:Deuteron-amp}
\end{align}
Comparing this to the dipole scattering amplitude \eqref{e:protondip1} on a single ``nucleon'' in the valence quark model, we obtain
\begin{align}  
 \Tr_C\,A^{q \bar q D} & \approx
  \left[ 2 \frac{\alpha_s}{N_c} \frac{1}{p_{1T}^{\prime 2}} \psi_{\sigma_D ; \, \sigma_p^\prime , -\sigma_p^\prime}^D (\ul{0} , \thalf)
  \delta_{\sigma_p^\prime , -\sigma_n^\prime} \right] \Tr_C\,A^{q \bar q N} ,
\end{align}
where in Eq.~\eqref{e:Deuteron-amp},  we have approximated the momentum fraction in the deuteron by its nonrelativistic value $\alpha = \thalf$ and used the momentum integral $d^2 p_1$ to Fourier transform the deuteron wave function to its value with zero transverse separation $\Delta x_\bot = 0$ between the nucleons.  From this relation, and using Eq.~\eqref{e:dipfact2} and Eq.~\eqref{e:dipfact4.1}), one obtains a relation between the gluon distributions\footnote{Note that in the approximations being considered here ($\alpha \approx \alpha' \approx \thalf$ and $p_{1T}^{\prime 2} \gg m_N^2$), the invariant $t$ defined by Eq.~\eqref{e:tdef} reduces to $t \approx - 2 p_{1T}^{\prime 2}$.}:
\begin{align} \label{e:Hfactoriz}
  \hat{H}^g_{(D)} (x_{eff}, 0, -\Delta_T^2; - 2 p_{1T}^{\prime 2}) \approx 
  \left[ 2 \frac{\alpha_s}{N_c} \frac{1}{p_{1T}^{\prime 2}} \psi_{\sigma_D ; \, \sigma_p^\prime , -\sigma_p^\prime}^D (\ul{0} , \thalf) \,
  \delta_{\sigma_p^\prime , -\sigma_n^\prime} \right] H^g_{(N)} (x_{eff}, 0, -\Delta_T^2) .
\end{align}
%
\begin{figure}[hbt]
 \centering
 \includegraphics[width= \textwidth]{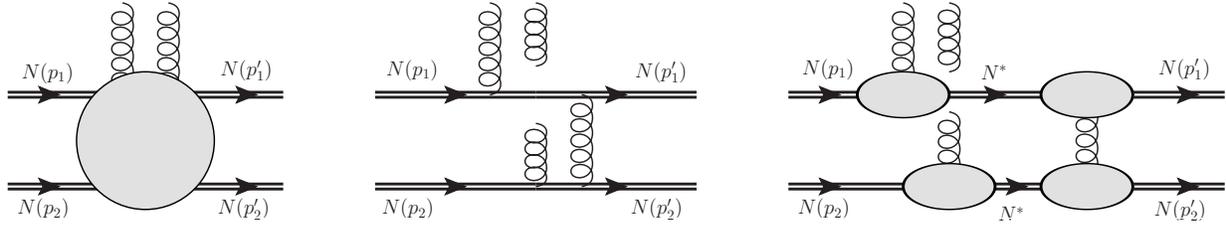}
 \caption{(Left Panel) Generic nonperturbative matrix element represented by Eqs.~\eqref{e:WFconv1} and \eqref{e:WFconv2}.  (Center Panel) Perturbative evaluation of the matrix element.
 (Right Panel) Possible factorization of the final-state rescattering based on the perturbative evaluation.}
 \label{f:GPD_Factoriz}
\end{figure}

This expression has a simple interpretation in terms of the proposed factorization illustrated in the right panel of Fig.~\ref{f:GPD_Factoriz}, which we write schematically as
\begin{align}
    \hat{H}_{(D)}^g \sim \psi^D (\ul{0}, \thalf) \times \left[ \frac{1}{N_c} H_{(N)}^g  \right] \times \left[ \frac{\alpha_s}{p_{1T}^{\prime 2}} \right] . 
\end{align}
The gluon field $\hat{H}^g_{(D)}$ of the deuteron seen by the projectile dipole is a simple product of the deuteron wave function at the origin ($\Delta x_T = 0$), the ordinary gluon distribution $H^g_{(N)}$ of the ``nucleons,'' and a new factor $\sim \alpha_s / p_{1T}^{\prime 2}$ describing the hard scattering between the ``nucleons'' in the final state.  

The additional suppression by the color factor $1/N_c$ occurs because the colors of  ``valence nucleons'' in our toy model cannot be chosen independently; the total color exchanged between the projectile and the composite $NN$ system must be color singlet in order to generate a rapidity gap; further, the total color acquired by any given nucleon must be individually color singlet. 

We also note that the simple factorization which arises here at leading order in the perturbative calculation is very far away from a demonstration of factorization to all orders.  However as we argued previously, there is a separation of characteristic time scales between the long time dynamics of the deuteron wave function, the instantaneous diffractive scattering with the projectile dipole and the final state interaction. This was manifest in the pinching of the poles of $\ell^-$ associated with the intermediate state between the diffractive scattering and final state interaction (Fig.~\ref{f:Poles_1}) which trap the intermediate ``nucleons" to be close to on-shell. Whether these features persist in the more general case requires a much more sophisticated calculation that we leave for future work.  

The relation \eqref{e:Hfactoriz} between the gluon distributions can be carried over to the cross-sections through the use of Eq.~\eqref{e:CS1} and Eq.~\eqref{e:CS2}, yielding for fixed values of the deuteron spin $\sigma_D$ and final state nucleon spins $\sigma_p^\prime , \sigma_n^\prime$
\begin{align}
  \left. \frac{d\sigma^D}{dT \, dt \, dy_1^\prime} \right|_{\alpha' \approx \thalf} =
  \left[ \frac{1}{(2\pi)^2} \frac{\alpha_s^2}{N_c^2} \frac{1}{p_{1T}^{\prime 4}} \left| \psi^D_{\sigma_D ; \, \sigma_p^\prime , -\sigma_p^\prime} (\ul{0},\thalf) \right|^2 \delta_{\sigma_p^\prime, - \sigma_n^\prime} \right]
  \times \frac{d\sigma^N}{dT} ,
\end{align}
where we have again emphasized the restriction $\alpha' \approx \alpha \approx \thalf$ and summed over the spins of the proton and neutron in the final state.  

In Appendix~\ref{sec:JerryWF}, we examine some common choices of the deuteron wave function used in phenomenology.  In these applications, the orbital part of the wave function is independent of the spin configuration, multiplying a separately normalized spin state.  Thus
\begin{align} \label{e:SOWF1}
	\psi^D_{\sigma_D ; \, \sigma_p \sigma_n} (\ul{0}, \thalf) = \psi_{orbit} (\ul{0}, \thalf)
	\braket{(\thalf, \tfrac{\sigma_p}{2}) \otimes (\thalf, \tfrac{\sigma_n}{2})}{(S_D, \sigma_D)},
\end{align}
where $\psi_{orbit}$ is the orbital wave function, and the bra-ket product represents a Clebsch-Gordan coefficient.  The ground state of the deuteron is in an $S$-wave orbital state, with spin quantum number $S_D = 1$, and the condition $\sigma_p = - \sigma_n$ arising from the scattering mechanism implies that there is only nonzero overlap with the spin state $\sigma_D = 0$.  When calculating the cross-section for unpolarized scattering, we should average over the deuteron spin and sum over the spins of the final state nucleons, to obtain 
\begin{align}
	\left. \frac{d\sigma^D}{dT \, dt \, dy_1^\prime} \right|_{\alpha' \approx \thalf} =
  \left[ \frac{1}{12 \pi} \frac{\alpha_s^2}{N_c^2} \frac{1}{p_{1T}^{\prime 4}} 
	\left| \Psi_D (\ul{0},\thalf) \right|^2 \right] \times \frac{d\sigma^N}{dT} .
\label{e:CS-toy}
\end{align}
Note that the wave function $| \Psi_D (\ul{0}, \thalf) |^2$ from Appendix A differs from $| \psi_{orbit} (\ul{0},\thalf)|^2$ by a factor of $\pi$ due to a different convention for the Fourier transform.  
With this result, and with input for the deuteron wave function and the measured diffractive cross-section on the nucleon at HERA, one can obtain the simplest possible perturbative estimate for the desired cross-section.  Quantitative results from this numerical analysis will be discussed in Section~\ref{sec:rates}.

\subsection{Models of the perturbative structure of final state nucleon-nucleon scattering}
\label{sec:qcount}
\begin{figure}[bt]
 \centering
 \includegraphics[width=0.8\textwidth]{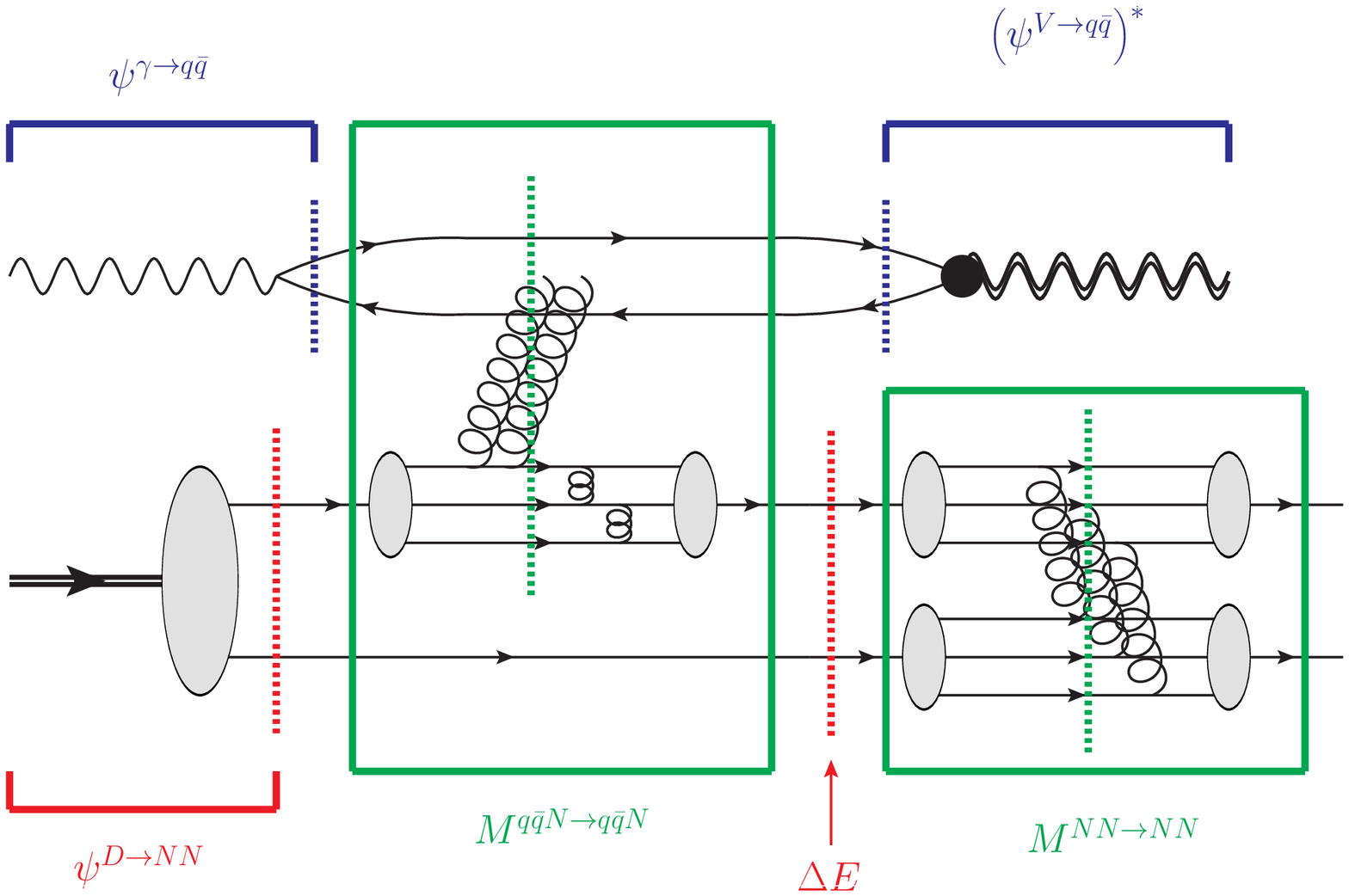}
 \caption{Breakdown of the amplitude for the process $\gamma^* D \longrightarrow J/\Psi N N$ with a hard color singlet exchange between the nucleons.}
\label{f:Singlet_NN}
\end{figure}

\begin{figure}[bt]
 \centering
 \includegraphics[width=0.8\textwidth]{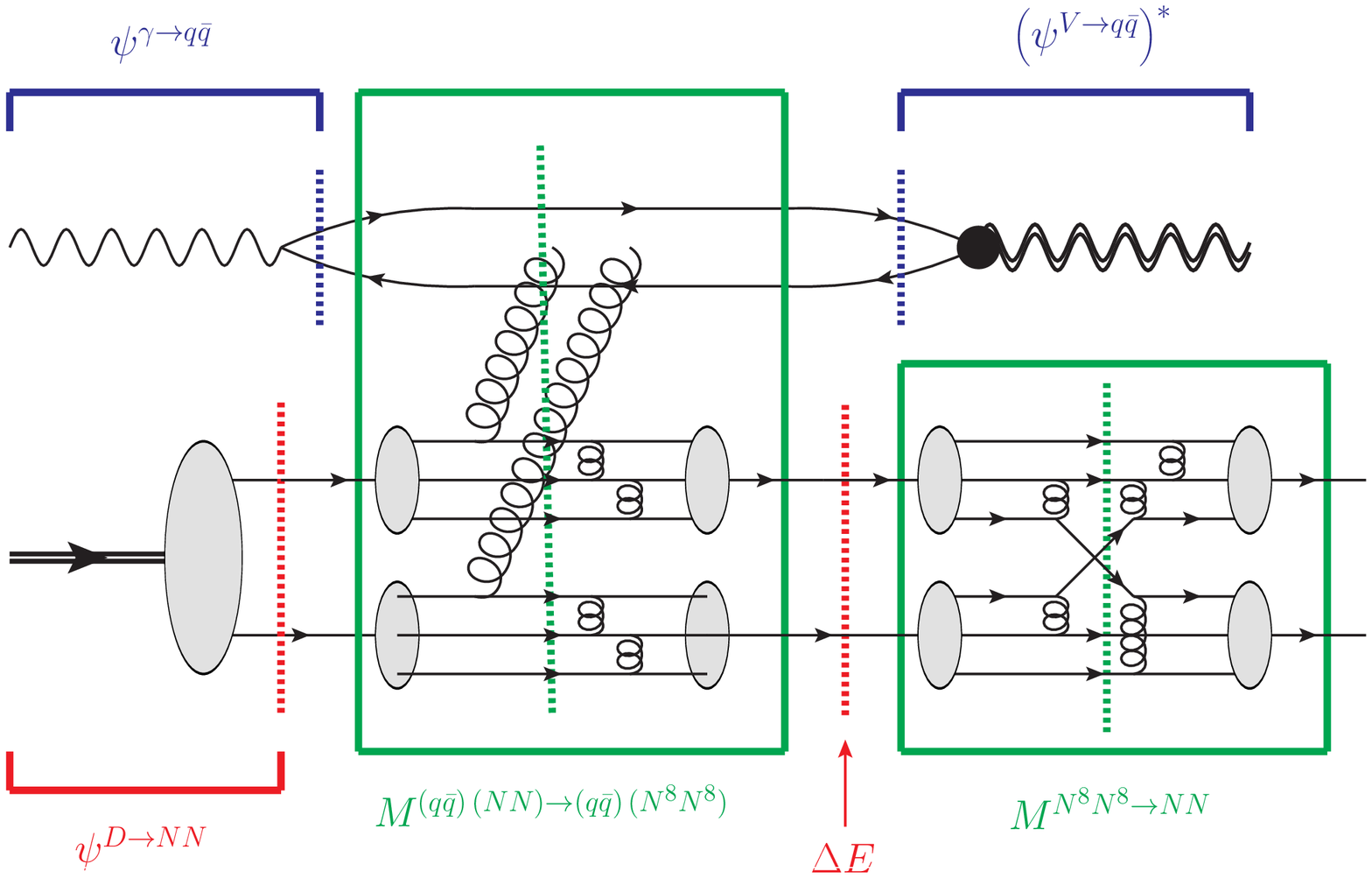}
 \caption{Breakdown of the amplitude the process $\gamma^* D \longrightarrow J/\Psi N N$ with a hard color octet exchange between the nucleons.}
\label{f:Octet_NN}
\end{figure}

In our discussion above, we considered a highly simplified model wherein the individual nucleons in the deuteron bound state are treated as valence quarks.  We performed a perturbative computation of DIS off these ``nucleons", with the exclusive production of a heavy vector meson (the $J/\Psi$) and back-to-back nucleons with high relative momenta in the final state. Though this model is not realistic, it does provide useful lessons and a first rough estimate of the rates for such a process. In a realistic computation, when $p_{1T}^{\prime 2} \gg m_N^2$, if nucleons are to remain collinear, additional gluon or quark exchanges must occur between the valence quarks in a nucleon and between the two nucleons. These additional exchanges break the naive power counting of the toy model which prefers a color octet final-state exchange; now both color singlet and color octet exchanges enter at the same parametric order.  Examples of such a process, with final state exchanges between the nucleons, are shown in Fig.~(\ref{f:Singlet_NN}) and (\ref{f:Octet_NN}). 

In the ``color singlet" exchange process shown in Fig.~(\ref{f:Singlet_NN}), the two gluons exchanged from the small sized dipole scatter off one of the nucleons. Ensuring that the three valence quarks remain collinear requires the outgoing nucleons to exchange multiple partons, an example of which is shown in the figure. On a much longer time scale relative to the gluon exchange from the projectile, a large transverse momentum $p_{1T}^\prime$ is transferred to the other nucleon via the color singlet exchange of three hard gluons--thereby ensuring that the valence quarks absorbing the gluons remain collinear after the scattering. 

In the language of high energy nucleon--nucleon scattering, such a process is called a Landshoff process~\cite{Landshoff:1974ew}.  The ``conventional'' mechanism shown in Fig.~(\ref{f:Singlet_NN}) is a color-singlet exchange, but a novel Landshoff process can occur if the two gluons from the projectile are absorbed on a different nucleon. In this case, the bound state is excited into a color octet-color octet configuration; they can then decay into the large relative momentum neutron-proton final state either by a color octet Landshoff mechanism or via a quark exchange diagram as shown in Fig.~(\ref{f:Octet_NN}). 

The discussion of high-energy, large-angle elastic scattering has a long history~\cite{Radyushkin:2002pe}. In pQCD, such scattering can asymptotically be represented as the product of the nonperturbative incoming and outgoing wavefunctions of the scattering nucleons, convoluted with a perturbative matrix element for the scattering~\cite{Efremov:1979qk,Lepage:1980fj}. 
In considerations of the latter, the quark counting rules describe configurations in which all the valence quarks of the incoming and outgoing nucleons are found within a small-sized ``color transparent" configuration~\cite{Bertsch:1981py}, so that they all participate in a single ``point-like'' hard scattering mechanism~\cite{Brodsky:1974vy,Matveev:1973ra}.  This leads to an elastic nucleon-nucleon scattering cross-section that goes as $d\sigma/dt \propto s^{-10}$.  Alternately, in the Landshoff or ``independent quark scattering" process, the valence quarks can scatter pairwise, without requiring that all the quarks be simultaneously within the same short-distance region.  This leads to a ``geometric enhancement'' of $s^2$ in the elastic scattering cross-section such that $d\sigma/dt \propto s^{-8}$.  The technical reason for this difference arises precisely from ``pinch singularities" of the sort we outlined in some detail in our toy model computation.

An example of these ``technical" contributions is the Sudakov suppression of soft gluon emissions~\cite{Sen:1981sd} which were conjectured to eliminate the Landshoff pinch singularities. However the Sudakov suppression does not eliminate the singularities, and only changes the power of $s$ in the cross-section~\cite{Mueller:1981sg} to one that is intermediate between the independent scattering and point-like processes. Indeed, the resulting infrared structure of the independent quark scattering model results in a ``Chromo-Coulomb" phase shift resulting from the interference between the long distance physics of the former and the short distance physics of the latter~\cite{Pire:1982iv}.  A remarkable consequence of this interference are oscillations in the energy dependence of fixed angle elastic scattering, which were observed experimentally in proton-proton collisions~\cite{Carroll:1988rp}. 

A systematic derivation of large angle nucleon-nucleon scattering, which includes the Landshoff pinch contributions as well as the noted effects due to the summation of Sudakov logarithms was initiated by Botts and Sterman~\cite{Botts:1989kf}. They studied the interplay between ``the geometric enhancement and radiative suppression" of the hard cross-section that controls the physics of hard large angle elastic scattering--an excellent review of the physics and subsequent developments from \cite{Botts:1989kf} onwards can be found in \cite{Sterman:2010jv}. (See also \cite{Botts:1990uy}.)

Much of the discussion thus far is based on high-energy asymptotics. It remains an open question whether this carries over to finite energies where experiments have been performed~\cite{Frankfurt:1994nn}. Another interesting avenue, which we have not discussed thus far, is the effect of spin dependence for polarized nucleon-nucleon scattering. In particular, for fixed target polarized proton-proton scattering at the BNL AGS, large spin-spin correlation asymmetries were observed at $s\approx 23$ GeV$^2$~\cite{Court:1986dh} which, along with the aforementioned oscillations, were alternatively interpreted as arising from the onset of exotic resonant structures in the vicinity of the open charm threshold~\cite{Brodsky:1987xw}. In summary, as discussed in \cite{Sargsian:2014bwa}, there are a number of issues that remain unresolved in first principles approaches to elastic nucleon-nucleon scattering at large momentum transfers. 

The process we identified therefore has the potential to cast fresh light on both the nonperturbative and perturbative aspects of short range nucleon-nucleon interactions as the relative $s$ between the measured photon and proton is varied. The ability to vary $Q^2$ as well as use a range of vector meson final states as probes allows us to search for this mechanism in a variety of processes to optimize the observable rates.  Further, our discussion of the deuteron can in principle be extended to ``knock-out" reactions--where back-to-back nucleons are produced in exclusive scattering off other light and heavy nuclei. 

It is therefore important to estimate the rates necessary to study this process, in particular the maximal accessible $p_T$'s.  A first principles perturbative computation, while feasible, is challenging and outside the scope of this work. However the take away lesson from our pQCD computation can be generalized to the process shown in Figs.~\ref{f:Singlet_NN} and \ref{f:Octet_NN}; we expect from the separation of time scales that the lower part of the amplitude factorizes into diffraction on a nucleon, times the LFPT energy denominator of the $NN$ state, times the $NN$ rescattering matrix element, as in Eq.~\eqref{e:Low4}.  
\footnote{We note that precisely because the presence or absence of pinches may be important, a full computation beyond our initial estimate will be necessary in future.}. 

One can therefore make the following ansatz:
\begin{align}
 \frac{d\sigma^{\gamma D \rightarrow V N N}}{dT dt dy} =  \frac{1}{(4\pi)^3} \frac{1}{(2 s)^2} 
 \left| M^{\gamma D \rightarrow V N N} (\Delta_T^2 , p_{1T}^{\prime 2}) \right|^2 \,,
\end{align}
with
\begin{align}
 M^{\gamma D \rightarrow V N N} (\Delta_T^2 , p_{1T}^{\prime 2}) &= \int \frac{d\alpha}{4\pi \alpha (1-\alpha)} \frac{d^2 p_1}{(2\pi)^2} 
 \psi^D_{\sigma_D ; \, \sigma_p^\prime \sigma_n^\prime} (\ul{p_1}, \alpha) \left[ 2 M^{\gamma N \rightarrow V N} (\Delta_T^2) \right]
 \notag \\ &\times
 \frac{1}{2 p^+ \Delta E^-} \left[ M^{N N \rightarrow N N} (p_{1T}^{\prime 2}) \right]
\end{align}
There is a factor of 2 at the amplitude level coming from diffraction proceeding on either the proton or neutron, and in principle there can be nontrivial spin dependence in the two amplitudes.  We assume that the scattering amplitudes are spin independent and helicity-conserving, which allows us to couple to any of the deuteron spin states.  Averaging over deuteron spins (including the sum over Clebsch-Gordan coefficients), summing over final-state nucleon spins, one obtains
\begin{align}
  \!\!\!\!\!\!
  \frac{d\sigma^{\gamma D \rightarrow V N N}}{dT dt dy} &=\frac{1}{64\pi^4} \frac{1}{(2 s)^2} \left| \Psi_D (\ul{0},\thalf) \right|^2
  \left| 2 M^{\gamma N \rightarrow V N} (\Delta_T^2) \right|^2 \frac{1}{(2 p^+ \Delta E^-)^2} \left| M^{N N \rightarrow N N} (p_{1T}^{\prime 2}) \right|^2
\end{align}
The energy denominator is
\begin{align}\label{e:sNN}
  2 p^+ \Delta E^- = 2 p^+ \left[ p_1^{\prime -} + (p-p_1^\prime + \Delta)^- - (p_1 + \Delta)^- - (p-p_1)^- \right]  \approx 4\,p_{1T}^{\prime 2} = s_{NN}  
\end{align}
assuming $\alpha^\prime = 1/2$, and the other amplitudes are related to the cross sections by $| M^{\gamma N \rightarrow V N} (\Delta_T^2)|^2 = 4\pi (s)^2\, \frac{d\sigma^{\gamma N \rightarrow V N}}{dT}$ and $| M^{N N \rightarrow N N} (s_{NN})|^2 = 4 \pi (2 s_{N N})^2 \frac{d\sigma^{N N \rightarrow N N}}{dT_{N N}}$. (Note that the photon-nucleon invariant mass is $\thalf s$.)  Putting it all together, one then obtains 
\begin{align}
  \frac{d\sigma^{\gamma D \rightarrow V N N}}{dT dt dy} &=\frac{1}{\pi^2} \left| \Psi_D (\ul{0},\thalf) \right|^2
  \left[ \frac{d\sigma^{\gamma N \rightarrow V N}}{dT} \right]_{T = - \Delta_T^2}
  \left[ \frac{d\sigma^{N N \rightarrow N N}}{dT_{N N}}  \right]_{T_{N N} = - p_{1T}^{\prime 2}}
\label{e:CS-realistic}
\end{align}
In the next Section, we will discuss the rates obtained from this cross-section at an EIC. 

\section{Estimation of Rates}
\label{sec:rates}
Diffractive $\Jpsi$ electro- and photo-production events were previously measured at HERA on a proton target for a range of energies corresponding to the photon-proton subprocess.  The analogous measurement of interest here is of small momentum transfer exchange $T\approx 0$ but with the target deuteron undergoing an additional interaction causing it to disintegrate with high $p_T$; this process will clearly be suppressed compared to the leading order process in which the target remains intact.  We will now address the question of what this margin of suppression is. Further by comparing with the kinematics from the HERA data, we will attempt to determine how readily accessible such a process should be at a future EIC.  

As a warmup, we will first do this estimate directly using the pQCD calculation of Section~\ref{sec:maincalc} in which the nucleons are treated as single valence quarks.  We will then  generalize this treatment to the more realistic case discussed in Section~\ref{sec:qcount}. The asymptotic power counting we will obtain for the former is likely very optimistic and may therefore be considered as an absolute upper bound to the rates one obtains with the latter at very high $p_T$.  In the toy example, if we integrate Eq.~(\ref{e:CS-toy}) over a finite window in $p_{1T}^{\prime}$ to implement a finite detector acceptance, and to restrict the kinematics to the regime of validity of our calculation, we obtain
\begin{align} \label{e:Rompot}
 \left. \frac{d\sigma^D}{dT \, dy_1^\prime} \right|_{\alpha' \approx \thalf} =
  \left[ \frac{1}{6 \pi} \frac{\alpha_s^2}{N_c^2} 
	\frac{\left| \Psi_D(\ul{0}, \thalf) \right|^2}{p_{T,min}^2}
  \left(1 - \frac{p_{T,min}^2}{p_{T,max}^2} \right) \right] \times \frac{d\sigma^N}{dT} .
\end{align}
The requirement that the invariant momentum transfer $|T| \approx \Delta_T^2$ to the $N N$ center of mass be small compared to the transverse momentum $p_{1T}^{\prime 2}$ of the final state nucleons and that $\Delta^- \approx \tfrac{s_{NN}}{s} q^-$ does not grow beyond the small-$x$ regime of Glauber gluon exchange gives the bound 
\begin{align}
	\mathrm{Max} \{ 1~GeV^2 , \tfrac{1}{4} |T| \} \ll p_{1T}^{\prime 2} \ll \frac{|T|}{4 x_{eff}}.
\end{align}
For $|T| \approx (\Lambda_{QCD})^2 \approx 0.04$ GeV$^2$, $Q = 0 , M_V \approx 3.1$ GeV and $s \approx (90~{\rm GeV})^2$ for an EIC, we obtain via Eq.~\eqref{e:xeff} that $x_{eff} \approx 1.2 \times 10^{-3}$.  This results in a window of validity of $1~{\rm GeV}^2 \leq p_{1T}^{\prime 2} \leq 8.4~{\rm GeV}^2$.

In addition to the theoretical limitations on the range of $p_{1T}^{\prime 2}$, we should also consider the practical limitations due to detector acceptance.  Simulations of Roman pot detectors at a future EIC -- performed with the eRHIC design specifications -- show that good acceptance of protons is feasible in a finite $p_T$ range $\approx 0.4~{\rm GeV} - 1.5~{\rm GeV}$~\cite{eicwikidvcs, Aschenauer:2014cki}.  This provides a more stringent upper bound of $p_{1T}^{\prime 2} \leq 2.25~{\rm GeV}^2$. Such events have also been discussed previously in the context of centrality selection in e+A collisions~\cite{Lappi:2014foa}; further precision in the measurement of the process of interest is allowed by installation of neutron detectors~\cite{Zheng:2014cha}. We note that neutron-proton coincidence studies have been performed previously at Jefferson lab--see for instance \cite{Korover:2014dma} and references therein. 

However since experiments at the EIC are a decade away, we will consider two scenarios: a ``Roman pot'' scenario suggested by the one extant simulation of a specific machine configuration and the ``Full'' scenario limited only by the constraints of theory. 

\begin{align} \label{e:Limits2}
	\mathrm{Roman~pot:}& \hspace{1cm} p_{T,min}^2 = 1~{\rm GeV}^2 \hspace{1cm} p_{T,max}^2 = 2.3~{\rm GeV}^2
	\notag \\
	\mathrm{Full:}& \hspace{1cm} p_{T,min}^2 = 1~{\rm GeV}^2 \hspace{1cm} p_{T,max}^2 = 8.4~{\rm GeV}^2 .
\end{align}

We will now like to use some realistic numbers in \eqref{e:Rompot} to estimate the cross-section for exclusive vector meson production with high-$p_T$ deuteron breakup.  For the coefficients, we take $\alpha_s \approx 0.3$ and $N_c = 3$, and we take the model value $\Psi_D (\ul{0},\thalf) \approx 1.05~fm^{-1}$ from \eqref{e:JerryWF} in Appendix \ref{sec:JerryWF} for the deuteron wave function.  We take the values of $p_{T,min}^2$ and $p_{T,max}^2$ from the kinematic windows of \eqref{e:Limits2}.  The last ingredient is the baseline diffractive cross-section on the nucleon.  For this, we take the ZEUS fit to $\Jpsi$ photoproduction data at HERA from \cite{Chekanov:2002xi}:
\begin{align} 
\label{e:Zfit}
 \left.\frac{d\sigma^N}{d|T|}\right|_{ZEUS} = \left(\tfrac{d\sigma^N}{d|T|}\right)_{T=0}  e^{- b |T|}\,\,;\,\,\left(\tfrac{d\sigma^N}{d|T|}\right)_{T=0} \approx 208 ~ nb / {\rm GeV}^2 \,\,;\,\, b \approx 4.02 ~{\rm  GeV}^{-2}.
\end{align}
It is instructive to use this fit to note separately the $p_{1T}^{\prime 2}$ dependence and the $|T|$ dependence of the cross-section.  In the fully differential case,
\begin{align}
	\left. \frac{d\sigma^D}{dT \, dt \, dy_1^\prime} \right|_{\alpha' \approx \thalf} =
  \left[ \frac{1}{12 \pi} \frac{\alpha_s^2}{N_c^2} \left| \Psi_D (\ul{0},\thalf) \right|^2 
	\left(\tfrac{d\sigma^N}{d|T|}\right)_{T=0} \right] \times
	\frac{1}{p_{1T}^{\prime 4}} \times e^{-b |T|} ,
\end{align}
and in the $p_T$-integrated case,
\begin{align}
	\left. \frac{d\sigma^D}{dT \, dy_1^\prime} \right|_{\alpha' \approx \thalf} &=
  \left[ \frac{1}{6 \pi} \frac{\alpha_s^2}{N_c^2} 
	\frac{\left| \Psi_D(\ul{0}, \thalf) \right|^2}{p_{T,min}^2}
  \left(1 - \frac{p_{T,min}^2}{p_{T,max}^2} \right) 
	\left(\frac{d\sigma^N}{d|T|}\right)_{T=0} \right] \times e^{-b |T|} .
\end{align}
%
\begin{figure}[hbt]
 \centering
 \includegraphics[width=0.7 \textwidth]{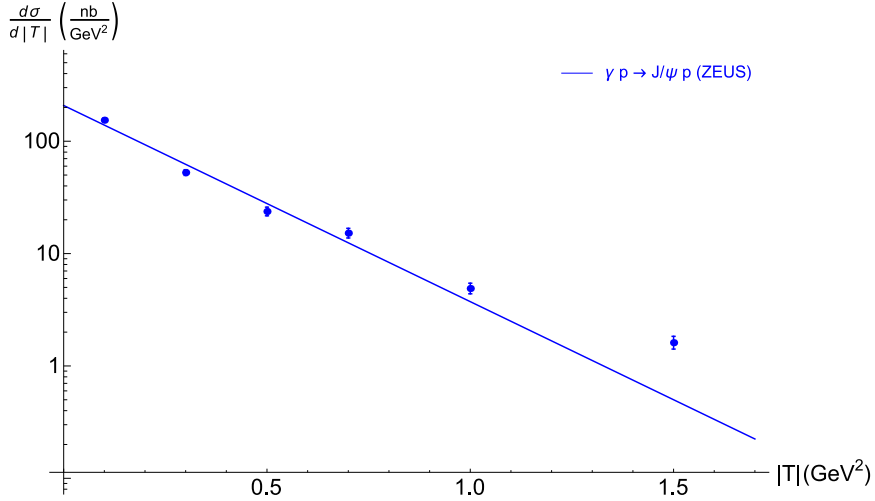}
 \caption{Selected data for $\Jpsi$ photoproduction at HERA.  The data and the fit \eqref{e:Zfit} are taken from Ref.~\cite{Chekanov:2002xi}, corresponding to a photon-proton center-of-mass energy $\sqrt{s} = W$ between $50$ GeV and $70$ GeV.}
 \label{f:ZEUS_Ratio} 
\end{figure}
Note that the $|T|$ dependence is exponential, reflecting the nonperturbative gluon distribution of the target, while the $p_{1T}^{\prime 2} \sim |t|$ dependence is a power law, reflecting the perturbative exchange of hard gluons\footnote{Dipole model frameworks, along the lines employed here, provide excellent fits to the combined ZEUS and HERA data for exclusive vector meson production~\cite{Rezaeian:2012ji}.}.  Using the values specified above, we obtain for the fully differential case
\begin{align} \label{e:ToyFit}
        \frac{d\sigma^D}{dT \, dt \, dy_1^\prime} &\approx
	\left[ 2.1 \frac{pb}{{\rm GeV}^4} \right] \left(\frac{1 ~ {\rm GeV}}{p_{1T}^{\prime}}\right)^4
\end{align}
and for the $p_T$-integrated case,
\begin{align}
	\frac{d\sigma^D}{dT \, dy_1^\prime} &\approx 
	\left[ 4.1~\frac{pb}{{\rm GeV}^2} \right] \left(\frac{1~{\rm GeV}}{p_{T,min}}\right)^2
	\left(1-\frac{p_{T,min}^2}{p_{T,max}^2}\right) \notag \\
	 & \approx
	\left[ 2.3 \, (3.7)~\frac{pb}{{\rm GeV}^2} \right] \hspace{0.3cm} \mathrm{Roman~pot \: (Full)} 
\end{align}

The ZEUS data corresponds to an integrated luminosity of only $38~ pb^{-1}$, which provided sufficient statistics out to $|T|=1.5$ GeV$^2$.  This is a useful baseline to estimate how easily the deuteron photodisintegration process could be measured at an EIC, by determining the integrated luminosity necessary to achieve comparable statistics to ZEUS' $|T|=1.5$ GeV$^2$ data point\footnote{We  note that this ZEUS data is taken for $50~GeV < W < 70~GeV$.}. Requiring the number of proton events $\mathcal{N}$ corresponding to an integrated luminosity of $\mathcal{L}_{ZEUS} = 38~pb^{-1}$ in the $|T|=1.5$ GeV$^2$ bin at ZEUS to be equal to the number of deuteron events at an EIC
\begin{align} \label{e:Lcrit}
  \mathcal{N} = \mathcal{L}_{ZEUS} \left[\left.\frac{d\sigma_p}{dT}\right|_{ZEUS}^{(1.5~{\rm GeV}^2)} 
	\right] \Delta|T|_{bin} &= \mathcal{L}_{EIC} \left[\left.\frac{d\sigma_D}{dT dy_1^\prime}\right|_{EIC} 
	\right] \Delta|T|_{bin} \Delta y_{bin}
\end{align}
corresponds to a necessary integrated luminosity for the EIC of
\begin{align}
	\mathcal{L}_{EIC} 
\approx
	\left[ 19~fb^{-1} \right] \left( \frac{ 1 ~ pb \, / \, {\rm GeV}^2} 
	{d\sigma^D \, / \, dT dy_1^\prime} \right)
	\approx \left[ 8.3\, (5.1) ~fb^{-1} \right] \hspace{0.3cm} \mathrm{Roman~pot}\, (\mathrm{Full}) ,
\end{align}
where we take $\Delta y_{bin} = 1~\mathrm{unit}$.  Given that the present EIC design goal for instantaneous luminosity is
\begin{align} \label{e:Lrate}
	(10^{33} - 10^{34}) ~ cm^{-2} ~ s^{-1} \sim (0.6 - 6) ~ fb^{-1} \, / \, wk ,
\end{align}
these statistics could potentially be achieved within about a week of EIC operation, assuming generous estimates for the luminosity. However as noted, this toy model calculation was done for ``nucleons'' consisting of single quarks so that all factors could be accounted for consistently.  The quark-quark scattering which drove the deuteron disintegration process is fairly weak, scaling as $\alpha_s / p_{1T}^{\prime 2}$ at the amplitude level.  
For more realistic treatment of the nucleons, the cross-sections at low transverse momentum are enhanced by nonperturbative form factors, while the cross-sections at large transverse momentum fall off much faster with $p_{1T}^\prime$ based on measurements of the $N N$ cross-sections.  

\begin{figure}[bt]
 \centering
 \includegraphics[width=0.7 \textwidth]{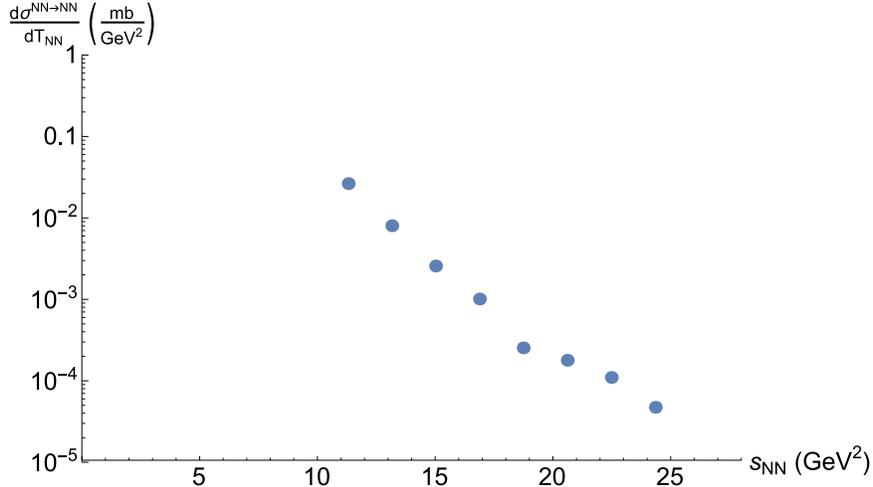}
 \caption{Proton-neutron elastic scattering cross-sections as a function of the nucleon-nucleon invariant mass $s_{NN}$.  The data points are interpolated to $\theta_{CM} = 60^\circ$ from the data tables of \cite{Stone:1977jh}.}
 \label{f:Stone_Xsec} 
\end{figure}

We will now use Eq.~(\ref{e:CS-realistic}) to estimate the rate. As previously, we will use $\Psi_D (\ul{0},\thalf) \approx 1.05~fm^{-1}$ from the appendix, and the ZEUS fit for exclusive photoproduction of $J/\Psi$ (for $|T| = 0.04 ~ {\rm GeV}^2$) which gives $d\sigma^{\gamma N} / dT = 177 ~ nb / {\rm GeV}^2$. For the neutron-proton cross-section, we will employ the available data on nucleon-nucleon scattering~\cite{Stone:1977jh}. The neutron-proton cross-section is shown in Fig.~\ref{f:Stone_Xsec} as a function of energy.  We note that the nucleon-nucleon cross-sections are commonly quoted for the center-of-mass scattering angle $\theta_{CM} = 90^\circ$; however, this does not quite correspond to the kinematics in our case.  The center-of-mass scattering angle is related to $s_{NN}$ and $T_{NN}$ by
\begin{align}
  T_{NN} = - (\thalf s _{NN}- 2 m_N^2) (1 - \cos\theta_{CM}),
\end{align}
where $s_{NN}$ is given by \eqref{sNN}, and $T_{NN} \approx - p_{T}^{\prime 2}$.  In the kinematics at hand, ($\alpha' \approx \thalf$ and $p_T^2 \gg m_N^2$) we have $T_{NN} \approx - \tfrac{1}{4} s_{NN}$, corresponding to $\theta_{CM} = 60^\circ$ rather than $90^\circ$.  Therefore we interpolate to $\cos \theta_{CM} = 0.5$ from the data tables of \cite{Stone:1977jh}, resulting in the cross-sections shown in Fig.~\ref{f:Stone_Xsec}.  The cross-sections for $\theta_{CM} = 60^\circ$ are about an order of magnitude larger than the ones for $\theta_{CM} = 90^\circ$.  As noted in \cite{Stone:1977jh}, the energy dependence of the neutron-proton cross-section at $60^\circ$ prefers a power law $\tfrac{d\sigma}{dT} \sim s_{NN}^{- 8.04}$, rather than the $s_{NN}^{-10.40}$ dependence shown by the $90^\circ$ data.  

\begin{figure}[bt]
 \centering
 \includegraphics[width=0.7 \textwidth]{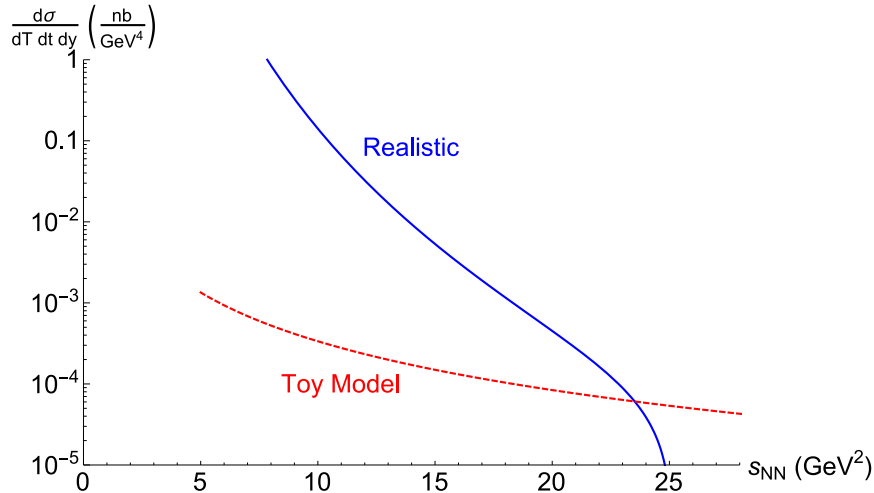}
 \caption{Cross-sections for $\Jpsi$ production with deuteron breakup using proton-neutron scattering data.  The valence-quark toy model (red dashed curve) of \eqref{e:ToyFit} has a smaller magnitude at low $s_{NN}$, but a slow decrease with energy as $s_{NN}^{-2}$.  The fit \eqref{e:CS-realistic} taken from NN scattering data (blue solid curve) has larger magnitude at low $s_{NN}$ due to nonperturbative form factors, but a very fast decrease with energy as $s_{NN}^{-8}$.}
 \label{f:Stone_Breakup} 
\end{figure}

\begin{figure}[bt]
 \centering
 \includegraphics[width=0.7 \textwidth]{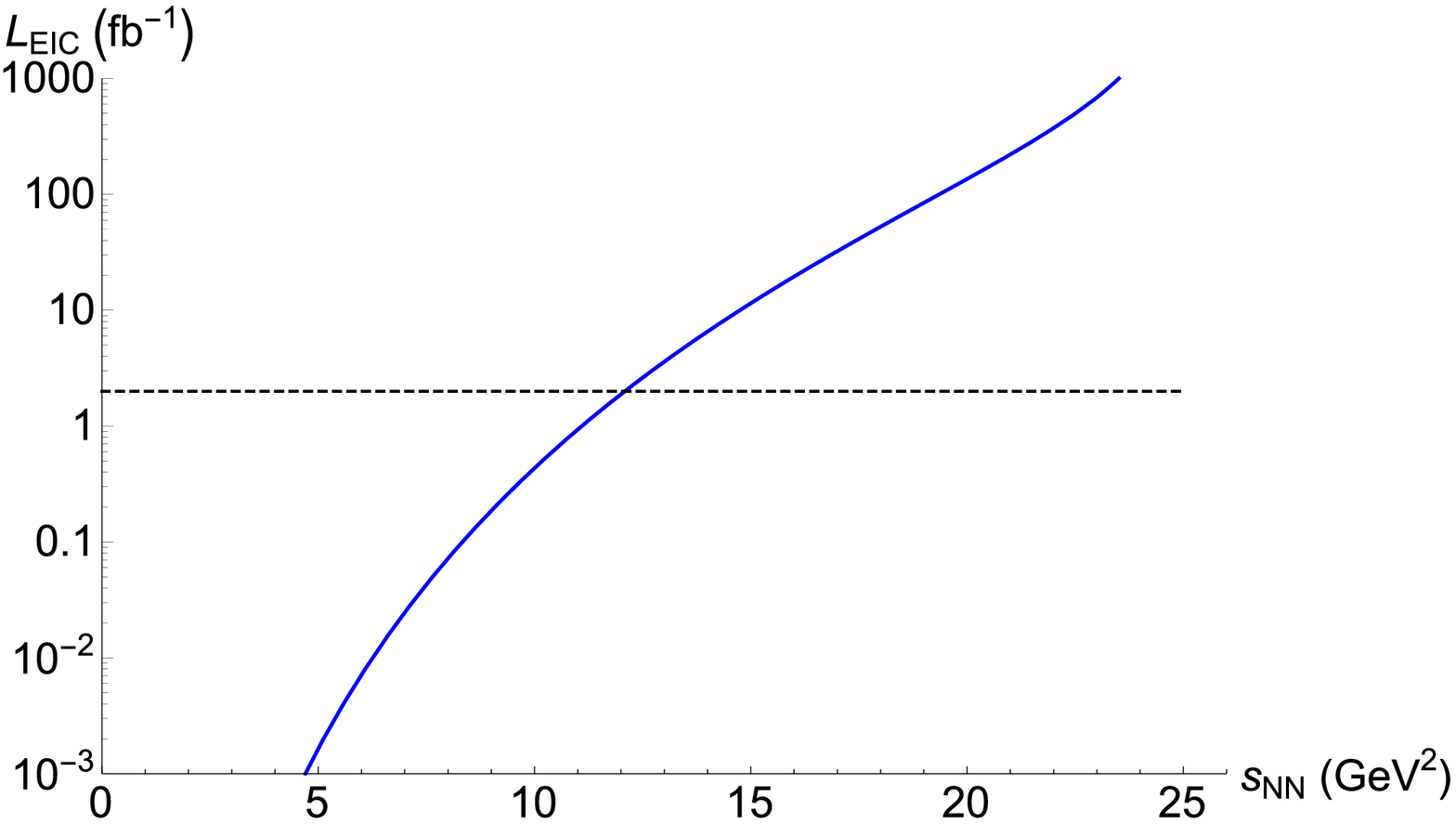}
 \caption{Integrated luminosities at an EIC needed to detect $J/\Psi$ production in the deuteron breakup process with a given $NN$ invariant mass $s_{NN}$.  With statistics from 20 weeks of running assuming a luminosity of $\sim 1.6\cdot 10^{32}$/cm$^2$/sec (black dashed line), a reasonable reach of $\sim 12$ GeV$^2$ can be obtained.}
 \label{f:Stone_Lumin} 
\end{figure}

Using the $NN$ cross-sections of Fig.~\ref{f:Stone_Xsec} in \eqref{e:CS-realistic}, we obtain the cross-sections for $\Jpsi$ photoproduction with hard deuteron breakup shown in Fig.~\ref{f:Stone_Breakup}.  Using the same criterion of Eq.~\eqref{e:Lcrit} to determine the necessary integrated luminosity at an EIC, we obtain the plot shown in Fig.~\ref{f:Stone_Lumin}.  A detection threshold of $2 ~ fb^{-1}$ is also shown, obtained from a conservative estimate in Eq.~\eqref{e:Lrate} of $0.1 ~fb^{-1}/wk$ (average luminosity of $1.6\cdot 10^{32}$/cm$^2$/sec) for the average instantaneous luminosity and assuming an operating time of 20 weeks out of the year.  The result, shown as the red line in Fig.~\ref{f:Stone_Lumin}, suggests that in this time frame an EIC could measure this breakup process on realistic nucleons out to $s_{NN} \sim 12$ GeV$^2$ with the same level of statistics obtained at HERA.  With more generous estimates of the EIC luminosity, this limit could extend out to $s_{NN} \sim 18 ~ GeV^2$.

%

\section{Summary}
\label{sec:Summary}
We considered here the process of back-to-back electro- or photo-production of high transverse momentum protons and neutrons accompanied by $J/\Psi$ production in DIS scattering off the deuteron. Such a process, which is a unique measurement at a future Electron-Ion Collider (EIC), offers the striking possibility of using  high energy quark-antiquark dipole pairs of varying size to probe short-distance correlations between protons and neutrons at much lower energies. To leading order, the nonperturbative matrix element characterizing the scattering, is a novel gluon Transition Generalized Parton Distribution (T-GPD), which is sensitive to the underlying structure of the short range nucleon-nucleon potential. Measurements of the T-GPD at an EIC can therefore significantly constrain models of the short range nuclear force. 

When the relative transverse momentum $p_T$ of the outgoing proton and neutron are large (with the center of mass energy of this 
subsystem expressed as $s_{NN}\sim 4 p_T^2$), perturbative computations of the scattering are feasible. We performed such a toy model computation with the nucleons replaced by valence quarks which demonstrated the importance of so-called ``pinch singularities" -- these ensure that the intermediate ``nucleon" states are close to being on-shell.  This permits a factorization, for large $s_{NN}$ asymptotics, of the T-GPD into a convolution of the deuteron wavefunction, the absorption of a color octet gluon by each of the intermediate nucleons, and final state octet gluon exchange between the two nucleons that ensures they remain color singlets. 

The full perturbative computation is quite challenging even at leading order. Nevertheless, it bears strong similarities to a vast literature on large angle elastic nucleon-nucleon scattering at high energies. The previous research addressed the relative importance of large sized Fock state configurations--the so-called Landshoff processes--where independent quarks in a nucleon each scatter off a partner in the oncoming nucleon, versus 
the importance of point like Fock configurations. In the latter, all the valence quarks participate in a single short-distance hard scattering. Quark counting rules devised for these processes give a different asymptotic $s_{NN}$ dependence from the Landshoff process.  Understanding these asymptotics, and extending our understanding to lower $s_{NN}$ is esential for uncovering the parton structure of 
short range nuclear forces. Further, as discussed,  the elastic scattering studies suggest a strong spin dependence to this parton structure, and the possible contribution of hitherto unobserved multi-parton configurations. 

Towards this end, exploiting the lessons from our toy model study and the extant literature, we made an ansatz that the cross-section for this process factorizes into a product of the squared deuteron waven function, the photoproduction cross-section for diffraction $J/\Psi$ production, and the neutron-proton elastic scattering cross-section. We note that our ansatz relied on a configuration where the two gluon exchange from the quark-antiquark dipole is off one of the nucleons, with a subsequent color-singlet exchange between the nucleons. As we discussed, other configurations are also feasible. With this ansatz, we were able to use information on the deuteron wavefunction (articulated in the appendix), data from HERA on exclusive photoproduction of $J/\Psi$ and data on neutron-proton elastic scattering cross-sections to estimate the rates for this process at an EIC. 

We imposed as a condition that the statistical accuracy of the data be comparable to that measured in exclusive photoproduction of $J/\Psi$ off the proton at HERA. Thus knowing the integrated luminosity of the HERA measurements, we were able to make estimates of the required luminosity at an EIC.  The toy model calculation, in which the deuteron T-GPD can be factorized as a product of the proton gluon distribution times a suppression factor $\alpha_S /(N_c \, {p^\prime}_T^2)$ arising from one gluon exchange in the final state, provides a lower bound for the rates at moderate $p_T^\prime$.  Even these small rates, however, are still accessible at an EIC with peak luminosities a factor of 10$^2$-10$^3$ that of HERA.  In the toy model, the cross-section falls off very slowly with increasing energy, scaling as $s_{NN}^{-2}$, which is likely overly optimistic.  

We showed that a more realistic ansatz of the exclusive photo-disintegration deuteron cross-section for $J/\Psi$ production can be expressed as a convolution of the deuteron wavefunction at the origin, the exclusive $J/\Psi$ photo-production cross-section at HERA and the neutron-proton $60^\circ$ elastic cross-section.  The rates for this channel are much larger at moderate $p_T^\prime$ due to nonperturbative form factors, but fall sharply with increasing energy as $s_{NN}^{-8}$.  These two models can be considered as upper and lower bounds for the behavior that can be expected at an EIC.  

Employing the known empirical values to compute the rates for the realistic ansatz, we find that for a conservative average luminosity of 
$1.6\times 10^{32} \, cm^{-2} \, s^{-1}$, comparable statistics to the HERA measurement can be attained with a 20 week run at an EIC for squared center-of-mass energies of the neutron-proton subsystem out to $s_{NN}\sim 12$ GeV$^2$. With these values of $s_{NN}$, it is feasible to scan the transition region from hadron to parton degrees of freedom in the description of short range nuclear forces. How the maximal $s_{NN}$ scales with luminosity will of course depend very sensitively on the power law dependence of the data with respect to this parameter. We stress that at present there are significant detector related challenges to extending this measurement out to even $s_{NN}=12$ GeV$^2$. 

The EIC offers the opportunity of extending the measurement outlined here to the quark-gluon dynamics of the target fragmentation region in DIS off polarized light nuclei as well as in large nuclei. In addition to photo-production of heavy vector mesons, the large $Q^2$ electro-production of light vector mesons will provide complementary insight. Recent experiments at Jlab, RHIC and the LHC that are sensitive to rare parton configurations in protons and nuclei have revealed a number of surprises; the measurement we have outlined, and like measurements, are important for a theoretical understanding of such states. Further, as our study clearly illustrates, the potential of EIC to scan the transition region from hadron to parton degrees of freedom at short distances, provides an important missing link of this physics to first principles studies of the structure of nuclei and neutron stars. 

\section*{Acknowledgements}

R. V. would like to thank Misak Sargsian and Mark Strikman for a number of insightful discussions over several years that stirred his interest in this topic. We would also like to thank Elke Aschenauer, Stan Brodsky, Rolf Ent, Dima Kharzeev, Tetsuo Hatsuda, Or Hen, Larry McLerran, Berndt Mueller, Jianwei Qiu, Sanjay Reddy and  Thomas Roser for useful comments. M. S and R.V. are supported under Department of Energy contract number Contract No. DE-SC0012704; the work of GAM  is partially supported by the U. S. Department of Energy Office of Science, Office of Nuclear Physics under Award Number DE-FG02-97ER-41014.  M. S. receives additional support from an EIC program development fund from BNL. R.V. would like to thank the Institut f\"{u}r Theoretische Physik, Heidelberg for their kind hospitality and the Excellence Initiative of Heidelberg University for their support.


\appendix



\section{Estimates of the Deuteron Wave Function}
\label{sec:JerryWF}


Here we discuss some model results for the wave function of the deuteron.  It is common in these nonrelativistic models to work with the convention
\bea \widetilde{\Psi}_D(\vec{p})=\int {d^3r\over (2\pi)^{3/2} }e^{-i \vec{p}\cdot\vec{r}} \: \Psi_D(\vec{r})\label{FT}\eea
and with both the coordinate- and momentum-space wave functions normalized to unity:
\bea 1 = \int d^3r \: \Psi_D^2(\vec{r}) = \int d^3p \: \widetilde{\Psi}^2_D(\vec{p}) . \eea 
In the nonrelativistic approximation, which is very accurate here, the momentum fraction is given by
\bea x={1\over2}+{p_z\over 2m},\eea
so that we can define the mixed-representation wave functions
\bea  \label{e:Jmixwf} \widetilde{\Psi}_D(p_\perp,x)=\sqrt{dp_z\over dx} \: \widetilde{\Psi}_D(p^2_\perp+ p_z^2(x))=\sqrt{2m} \: \widetilde{\Psi}_D(p^2_\perp+ m^2(2x-1)^2), \eea
where we have taken the wave functions to be spherically symmetric ($S$-wave).  With the normalization \eqref{e:Jmixwf}, the mixed wave functions are normalized as 
\bea1 \label{e:JerryNorm1} =\int d^2p_\perp\int\,dx\, \widetilde{\Psi}^2_D(p_\perp,x) = \int d^2r_\perp\int\,dx\, \Psi^2_D(r_\perp,x) .\eea

First let us consider the simple model of \cite{Tiburzi:2000je}:
\bea \widetilde{\Psi}_D(p_\perp,x)={1\over \pi}\sqrt{2m}\sqrt{a b(a+b)\over (a-b)^2}\left[{1\over  a^2 +m^2(2x-1)^2+p_\perp^2}-{ 1\over b^2+m^2(2x-1)^2+p_\perp^2}\right]\eea
where $a=$ 0.23161 fm$^{-1}$, $b=$ 1.308 fm$^{-1}$, $m$= nucleon mass =4.76 fm$^{-1}$.  We want the wave function at zero transverse separation,
\bea \label{e:JerryFT1} \Psi_D(\underline{r}=0,x)=\int {d^2p_\perp\over 2\pi}\,\widetilde{\Psi}_D(p_\perp,x), \eea
and direct integration gives
\bea \Psi_D(\underline{r}=0,x)={1\over \pi}\sqrt{2m}\sqrt{a b(a+b)\over (a-b)^2}{\pi\over 2\pi}  \log{b^2 +m^2(2x-1)\over  a^2+m^2(2x-1)}\eea
We can see this is narrowly peaked about $x=1/2$ (but not as narrowly peaked as more realistic wave functions.
Evaluation at $x=1/2$ gives
\bea \Psi_D(\underline{r}=0,x=1/2)={1\over \pi}\sqrt{m\over 2}\sqrt{a b(a+b)\over (a-b)^2}\log{b^2\over a^2}\label{main}\eea
so that
 \bea \Psi_D(\underline{r}=0,x=1/2)=1.1 {\rm fm}^{-1}=0.22 \,{\rm GeV}\eea
 
It is also convenient to write the wave function at $\ul{r}=0 , x= \thalf$ in terms of the spherically-symmetric coordinate-space wave function:
\bea& \Psi_D(\underline{r}=0,x=1/2)=\sqrt{2m}\int {d^2p_\perp\over 2\pi}\int {d^2r_\perp\over (2\pi)^{3/2}}e^{-i \vec{p}_\perp\cdot\vec{r}_\perp}\int_{-\infty}^\infty dz\,\Psi_D(\sqrt{r_\perp^2+z^2})\\&
=\sqrt{m\over \pi}\int_{-\infty}^\infty dz\,\Psi_D(\sqrt{z^2})
\eea
This last expression has a nice physical interpretation: fixing $p_z=0$ means all values of $z$ contribute.
Using instead the ANL V18 potential \cite{Wiringa:1994wb} gives 
\bea& \Psi_D^{AV18}(\underline{r}=0,x=1/2)=1.01\, {\rm fm}^{-1} . \eea
Similarly the Reid93 potential~ \cite{Stoks:1994wp}gives
\bea& \Psi_D^{R93}(\underline{r}=0,x=1/2)=1.05\, {\rm fm}^{-1}, \eea
and the Nimeggen II  potential  \cite{Stoks:1994wp} gives
\bea \label{e:JerryWF}  & \Psi_D^{N}(\underline{r}=0,x=1/2)=1.05\, {\rm fm}^{-1}.\eea
All three of these wave functions contain a 6\% D-wave state as well, so these numbers could be multiplied by 1.06 to compensate.

\providecommand{\href}[2]{#2}\begingroup\raggedright\endgroup


\end{document}